# State of Understanding of High-speed Combustion: Can the Detonation Structure in Fuel-Air Mixtures Be Predicted?

Alexei Y. Poludnenko[a,b], Sai Sandeep Dammati[a,b], Abeetath Ghosh[a]

[a]School of Mechanical, Aerospace, and Manufacturing Engineering, University of Connecticut, CT, USA - 06269
[b]Center for Clean Energy Engineering, University of Connecticut, Storrs, CT, USA - 06269

**1) Novelty Statement**

This work presents the first systematic, uniform survey of the multi-dimensional premixed detonation structure across a wide range of practically relevant fuels, namely hydrogen, methane, acetylene, ethylene, JP-10, and *n*-dodecane in air, under atmospheric conditions, as well as at elevated pressures and temperatures, using high-fidelity 2D and 3D simulations with the state-of-the-art multi-step kinetics and rigorously enforced numerical criteria. The study establishes for the first time in a single consistent framework that the disagreement between numerical and experimentally measured detonation cell sizes is systematic, it correlates with the thermochemical state space occupied by the detonation front rather than with pressure or numerical resolution, and it coincides with the high-pressure, high-temperature regimes, in which direct ignition-delay measurements in air-like compositions are limited.

**2) Significance Statement**

Detonation-based hypersonic propulsion, space flight, industrial explosion safety, and astrophysical thermonuclear Type Ia supernovae explosions all depend on our ability to predict detonation properties. Yet this work demonstrates that state-of-the-art numerical detonation models fail to do so for realistic fuels at practically relevant pressures, with characteristic detonation cell sizes under-predicted by factors 3-6 or more. By systematically connecting these observations with various aspects of the physicochemical models in use today to study strongly shocked, high-pressure conditions, and by cataloging where direct experimental validation data exist and where they do not, this paper suggests necessary improvements to the future experimental measurements, non-equilibrium kinetic models, and computational advances required to make high-speed combustion predictive.

**3) Author Contribution**

- AYP : research conceptualization, wrote and reviewed manuscript, data analysis, funding acquisition.
- SSD : solver development, simulations, data analysis, reviewed and edited manuscript.
- AG : simulations, data analysis.

# State of Understanding of High-speed Combustion: Can the Detonation Structure in Fuel-Air Mixtures Be Predicted?

Alexei Y. Poludnenko[a,b,*], Sai Sandeep Dammati[a,b], Abeetath Ghosh[a]

[a] School of Mechanical, Aerospace, and Manufacturing Engineering, University of Connecticut, Storrs, CT, USA - 06269
[b] Center for Clean Energy Engineering, University of Connecticut, Storrs, CT, USA - 06269

**Abstract**

High-speed reacting flows are central to many natural and practical systems, from detonation-based propulsion to industrial safety to various explosive astrophysical phenomena. Here we discuss the limitations in our understanding of high-speed combustion by asking how well modern state-of-the-art physicochemical and computational models can predict the properties of a quintessential, classical example of high-speed reacting flows - a freely propagating detonation. We present a systematic survey of eleven fuel/air mixtures, from hydrogen through methane, acetylene, and ethylene to complex JP-10 and *n*-dodecane jet fuels, at a range of upstream pressures, temperatures, and equivalence ratios. Detonation properties are analyzed using prior numerical studies in the literature, as well as the multi-dimensional, high-fidelity simulations performed in this work, which employ modern multi-step kinetics based on the FFCM/HyChem families of reaction mechanisms. We show that simulations reproduce experimental cell sizes in $H_2$/air at sub-atmospheric pressures or elevated upstream temperatures, yet at the atmospheric conditions, disagreement in the predicted cell sizes is a factor of 3 - 6 at stoichiometric composition and much more severe in lean mixtures. Hydrocarbon fuels show comparable or even larger discrepancies. We argue that such disagreement cannot be ascribed solely to the computational factors, in particular, numerical resolution even at the continuum limit. We analyze the thermodynamic state space occupied by detonations in various mixtures, describe the predicted chemical properties in various regions of the state space, in particular ignition delays and their temperature sensitivity, and consider how well modern chemical kinetics mechanisms are constrained at these conditions by the direct ignition delay measurements. We further assess the proposed role of vibrational non-equilibrium and demonstrate that the current evidence is inconclusive. Finally, we summarize key open questions, as well as the experimental, theoretical, and computational advances required to make the high-speed combustion fully predictive.

*Corresponding author.

## 1. Introduction

A wide range of practical, as well as natural systems involve reacting flows, which can be considered high-speed. Examples include detonation-based engines [1], hypersonic propulsors (scramjets) [2], industrial and natural accidents [3], as well as a variety of astrophysical phenomena, including explosive thermonuclear burning found in type Ia [4] and core-collapse supernovae [5], X-ray flashes [6], etc. All such systems and phenomena can appear qualitatively different as they involve very different media, i.e., gases or degenerate, relativistic plasmas, different physical processes, namely chemical or thermonuclear reactions, and different coupling with the ambient flow, to name a few. At the same time, they all share a common natural characteristic - fast convective velocities, and thus high Mach numbers, which lead to the formation of strong shocks. Therefore, for the purposes of this discussion, we define high-speed combustion as the exothermic physicochemical process that occurs in a shocked flow. It would be impossible to describe in detail both chemical and thermonuclear systems, which involve very different but equally fascinating physics. Hence, here we will focus on the chemical gaseous flows present on Earth.

The physics of high-speed reacting flows is both diverse and complex. An extremely wide range of thermodynamic conditions, including high pressures and temperatures, which can form in such flows due to the strong compressibility, can give rise to physical pro-cesses which may not be present in more traditional low-speed combustion systems. These include fast chemistry, real-gas effects, and pronounced thermal non-equilibrium formed behind strong shocks, which are all closely coupled creating complex dynamical interplay. To make matters worse, these diverse physical processes invariably occur in the presence of fast turbulence, the multi-scale and highly non-linear nature of which adds an additional, and formidable, layer of complexity. Such turbulence can drive rapid mixing and, at the same time, create large gradients of both active and passive scalars. This results in an extremely wide range of spatial scales, which can extend from the system size down to the molecular mean free path, and temporal scales, which can approach nanoseconds. Such multi-scale and multi-physics nature of high-speed combustion challenges our ability to study it from first principles both theoretically and in high-fidelity numerical computations.

It would be impossible to discuss in any meaningful detail in the limited space of this paper all physical aspects of the various high-speed reacting flows that arise in different practical settings. Instead, we attempt to answer one central question: what is the state of our understanding of high-speed combustion? To this end, we focus our discussion on a quintessential, classical example of a high-speed reacting flow - a freely propagating detonation.

This shock-driven reaction front has been actively investigated for more than a century [7–9]. Despite the considerable experimental and theoretical insights obtained over that time, detonation studies still remain an active and vibrant area of combustion science, with many unsolved questions central to this phenomenon [10, 11]. Canonical gaseous detonation represents one of the most extreme flows, which can exist on Earth. It is difficult to find another system, which does not involve nuclear processes, with similar characteristics: remarkable power densities, which can reach gigawatts per square meter of the front, shock strengths, which are otherwise found only in hypersonic systems and reentry vehicles with Mach numbers as high as $\sim 10$ and above, pressures exceeding thousands of atmospheres and temperatures reaching $4000 - 5000$ K. Such remarkable properties do not require exotic and complex experimental setups, as they naturally exist in a detonation freely propagating at the relatively mundane upstream conditions with pressures and temperatures close to, or somewhat higher than, atmospheric. Thus, a detonation can be viewed as a unique physical laboratory, which allows us to directly access extreme flow regimes, and therefore assess how well we understand various aspects of high-speed combustion, from the thermophysics and chemical kinetics to numerical modeling.

In the discussion below, we do not consider many of the important and fascinating dynamical aspects of a detonation, for instance, mechanisms of detonation ignition, including the deflagration-to-detonation transition [12, 13], or the detonation quenching and near-limit regimes [14, 15]. Instead, we focus on the question of how well modern state-of-the-art physicochemical and computational models can predict the properties, structure, and most importantly internal dynamics of a multi-dimensional detonation. We specifically attempt to answer this question in the context of realistic and practically relevant fuels – from hydrogen to methane to jet fuels used in the propulsion and energy conversion systems. We focus on fuel-air mixtures, which find their applications in air-breathing propulsion and power generation, as well as various industrial settings. Mixtures in pure $O_2$, which could be relevant, for instance, to detonative rocket propulsion, deserve a detailed review in their own right, which is outside the scope of this paper. We also do not consider mixtures with argon or other inert diluents, which are primarily of academic interest. Finally, we restrict the discussion to the simplest case of a planar, freely propagating detonation, and thus we do not consider more applied aspects related to the actual engineering systems, including fuel injection, mixing, etc.

In the context of detonations in the simplest and arguably best studied combustion fuel, hydrogen, there exist a considerable number of studies, which have analyzed detonation properties in fuel-air mixtures using complex, multi-step chemical mechanisms both at sub-atmospheric [16, 17] and atmospheric pressures [18–20] (see Table 1 and Table S1 in the Supplementary Material for the detailed summary of prior numerical studies). It is worth noting that at lower pressures, the vast majority of studies were performed in pure $O_2$ or Ar diluted mixtures [21–23], which are outside the scope of this discussion.

At atmospheric conditions, several earlier studies have observed discrepancies in the detonation structure between the numerical models and experiments [18–20]. In particular, detonation cells in a stoichiometric $H_2$/air mixture were found to be a factor of $\approx 3 - 6$ smaller than the experimental values, which is greater than the potential experimental uncertainties (Table 1). To the best of our knowledge, no systematic survey has been conducted for a wide range of hydrogen mixtures and thermodynamic conditions to establish the nature of such a disagreement. As a result, there is no clear understanding whether such a disagreement indeed exists, what its extent is, and most importantly what the causes of it are. Over the years, various potential sources have been proposed, from numerical in nature, e.g., lack of resolution [24], to the limitations of the physical model, including the breakdown of the Navier-Stokes model [25] or the lack of treatment of the vibrational non-equilibrium [26–28].

For hydrocarbon fuels, both light and heavy, the situation is even worse as there are far fewer numerical studies with detailed kinetics than in hydrogen [14, 29–32], with most of those prior works considering mixtures either in pure $O_2$, with Ar dilution, or with $N_2$ but with significant $O_2$ enrichment. Thus, it is even less clear whether existing computational models are able to accurately predict the detonation properties in hydrocarbons.

Our primary objective is to perform a systematic survey of a wide range of mixtures, from hydrogen to hydrocarbons, and upstream conditions (pressures and temperatures) using a uniform and well characterized numerical approach and physicochemical model. The goal is to contrast the predictions of modern, multi-dimensional, high-fidelity simulations of detonations with the experimental observations of the detonation structure, in particular a detonation cell size.

The discussion presented here, by no means, claims to provide the final answer. Instead, our hope is to take the first realistic and comprehensive view of the fidelity of modern physicochemical and numerical models of high-speed combustion under practically relevant conditions. As will be argued in the following, there exist, in our view, major gaps in the understanding of high-speed combustion, which severely limit the predictive ability of modern theoretical and computational tools. With this contribution, we aim to start the discussion in the community of these limitations in the theory of such a critically important and under-explored combustion regime in order to identify the potential path forward, which could remedy these shortcomings.

## 2. A Few Considerations Regarding Numerical Simulations

A numerical simulation can provide meaningful information regarding the intrinsic detonation properties if several necessary conditions are satisfied:

- **Detonation speed** must be relaxed to $\lesssim 1\%$ of the speed of a Chapman-Jouguet (CJ) detonation, $D_{CJ}$. This most basic requirement is merely the reflection of the conservation properties of a numerical solver, and thus it is not sufficient to ensure an accurate overall numerical solution.
- **Initial conditions** must not imprint a particular spatial scale on a detonation. This requirement is less critical in irregular mixtures. However, in detonations producing a regular cellular structure, cell size can be imposed by a particular method used to initialize the solution in the domain [33, 34].
- **Channel width** must be sufficiently large to prevent mode-locking [35]. Typically, it must be $\gtrsim$ 4-5$\lambda_{exp}$, where $\lambda_{exp}$ is the experimental cell size. Channel sizes of $1-2$ cells, or less, often used in simulations to reduce the computational cost, are not sufficient to allow the front to develop its natural structure.
- **Temporal convergence.** Finally, and most importantly, a detonation must evolve over a sufficiently long distance to ensure that the cellular structure also reaches its natural state and becomes temporally converged. In particular, it is critical to show that the cellular structure remains statistically steady over a distance of at least 2-3 channel widths, and preferably more.

The last condition is especially important and is often overlooked, which can lead to erroneous conclusions. A detonation is typically formed in the overdriven state, and its speed may relax to $D_{CJ}$ faster than its overall structure. Furthermore, in the process of this relaxation, characteristic detonation cell size often evolves non-monotonically, especially in highly irregular, unstable mixtures in two dimensions (2D). Therefore, rigorous demonstration of the temporal convergence of the detonation structure is one of the key characteristics of a robust numerical result.

Over the years, a significant number of studies reported numerical simulations of different degree of fidelity and realism, which targeted various aspects of the detonation dynamics. Of that body of work, over 60 papers to-date have studied what can be qualified as a freely propagating detonation in a variety of fuel compositions and upstream conditions using a realistically complex, multi-step combustion chemistry, which was not specifically tuned to reproduce some specific detonation characteristics. In Table S1 in Supplementary Material we summarize a subset of those computational studies, which targeted fuel-air (or near-air) compositions and which provide information regarding the detonation cellular structure in the form of either, preferably, soot foils or specific values of the cell sizes (or both). We note that at sub-atmospheric pressures, the vast majority of numerical studies have considered either pure $O_2$ mixtures, or mixtures with Ar dilution, rather than air. At atmospheric and higher pressures, air, or near-air, compositions are much more prevalent.

Crucially, not all studies listed in Table S1 (Supplementary Materials) allow robust comparison with the experimentally measured detonation structure for failing to satisfy the requirements listed above. In particular, many of the prior simulations were performed in narrow channels only one or two detonation cells wide (or sometimes even less), which could result in a mode-locked structure not representative of a freely propagating detonation. This is especially the case for prior three-dimensional (3D) simulations of $H_2$/air detonations [36–38]. Also, few studies provide evidence of the independence of the obtained detonation structure on the initial conditions or report the detonation speed and demonstrate temporal convergence of the soot foils, which would show that a detonation has reached a steady state. Finally, for some cases, no experimentally measured cell sizes are available.

Given these constraints, in Table 1, we list representative numerical studies, which provide a quantitative comparison of the predicted detonation structure with the experimental observations. Note that if for the same mixture conditions, multiple simulations were reported in the same study, for instance, which solved Navier-Stokes and Euler equations, used several chemical reaction mechanisms, or considered different channel widths, Table 1 lists the range of the reported cell sizes.

Table 1 shows that there is a rather limited number of studies, which can provide robust insights into the

Table 1: Representative numerical studies, which report detonation cell sizes in fuel/air (or near-air) mixtures. All studies listed were performed in a 2D geometry. $H$ is the channel width, $\lambda_c$ is the reported numerical cell width, and $\lambda_{\text{exp}}$ is the experimental cell width for the same mixture (cf. Table 2). See Table S1 in Supplementary Material for further details and additional studies.

| Ref. | Mixture | $P$ (bar) | $T$ (K) | $\phi$ | Model | $L$ (cm) | Soot foil? | $\lambda_c$ (cm) | $L/\lambda_c$ | $\lambda_c/\lambda_{exp}$ |
|---|---|---|---|---|---|---|---|---|---|---|
| [17] | $H_2$/air | 0.25 | 300 | 1.0 | NS | 3.77–5.53 | No$^\dagger$ | 0.45–1.09 | 5.07–8.38 | 0.19–0.45 |
| [16] | $H_2$/air | 0.40 | 300 | 1.0 | NS | 5.79 | Yes | 1.19 | 4.87 | 0.91 |
| [18] | $H_2$/air | 1.0 | 293 | 1.0 | Euler | 4.0 | Yes | 0.3–0.4 | 10.0–13.33 | 0.25–0.33 |
| [19] | $H_2$/air$^\ddagger$ | 1.0 | 500 | 1.0 | Euler | 5.0 | No | 0.31–0.67 | 7.46–16.13 | 0.49–1.06 |
| [20] | $H_2$/air$^\ddagger$ | 1.0 | 300 | 1.0 | Euler | 5.0 | Yes | 0.2 | 25.0 | 0.17 |
| " | $H_2$/air$^\ddagger$ | 1.0 | 500 | 1.0 | Euler | 5.0 | Yes | 0.31–0.7 | 7.14–16.13 | 0.49–1.11 |
| " | $H_2$/air$^\ddagger$ | 1.0 | 650 | 1.0 | Euler | 5.0 | Yes | 0.5 | 10.0 | 1.11 |
| [31] | $CH_4/O_2/N_2$$^\P$ | 1.0 | 298 | 1.0 | NS | 6.0 | Yes | 1.22 | 4.92 | 1.14 |
| " | $C_2H_2/O_2/N_2$$^\P$ | 1.0 | 298 | 0.4 | NS | 2.0 | Yes | 0.96 | 2.08 | 0.70 |
| [32] | JP-10/air | 1.0 | 373$^*$ | 1.0 | Euler | 25.0 | Yes | 5.0$^\|$ | 5.0 | 0.96 |

$^\dagger$The cells are shown as maps of $D/D_{\text{CJ}}$. $^\ddagger$The composition is close to, but not exactly, air. $^\P$$O_2$-enriched air. $^*$ Simulations performed at 373 K; experiments used for comparison performed at 353 K. $^\|$ Detonation has not reached a steady state or no temporal convergence of the cellular structure.

existing state of understanding of the realistic detonations, and of high-speed combustion in general. To address this shortcoming, we performed a series of high-fidelity simulations in a wide range of fuel-air mixtures, from $H_2$ to heavy hydrocarbons, which we describe below.

Among the conditions above, we do not list grid convergence, demonstration of which must be the starting point of any rigorous numerical study. The main difficulty here is that achieving such convergence often is not possible in realistic fuels, even when using state-of-the-art numerical methods and multi-species thermochemical models. Although convergence has been demonstrated in mixtures that produce regular detonations [39, 40], in irregular mixtures, grid convergence may not be achieved even when the computational grid cell size is reduced to the continuum, or slip flow, limit with the Knudsen number $Kn \sim 0.1$. This was first shown by Taylor et al. [18], and we discuss this issue further below. At the same time, for consistency with previous numerical studies [16, 40], we adopt the following two resolution criteria for all simulations performed by us: (i) $\Delta x \leq \Delta_i/40$ in the 2D cases and $\Delta x \leq \Delta_i/20$ cells in the 3D calculations, and (ii) $\Delta x \leq \Delta_e/5$ both in 2D and 3D. Here, $\Delta_i$ and $\Delta_e$ are, respectively, the induction length and exothermic widths in the corresponding Zel'dovich-von Neumann-Döring (ZND) solution. Finally, in all calculations discussed here, detonation was allowed to propagate at least ten channel widths, $L$, to fully achieve a steady state, although in some cases achieving temporal convergence required propagation distances of $20-40L$.

Numerical simulations that we describe here were performed using the code `Athena-RFX++` [59, 60], which evolved from the code `Athena-RFX` used extensively over the years to study chemical and thermonuclear reacting flows [13, 16, 61–67]. `Athena-RFX++` re-implements the numerical algorithms of `Athena-RFX` in a modern, template-based C++ framework with a new data structure layout, which enables 99% code vectorization. The code employs several fully compressible, finite-volume solvers based on the Van Leer and various-order Runge-Kutta unsplit schemes [60, 68], along with the piecewise-linear (PLM) and piecewise-parabolic (PPM) spatial state reconstructions, which together provide 2$^{\text{nd}}$- and 3$^{\text{rd}}$-order accuracy in space and 2$^{\text{nd}}$ to 4$^{\text{th}}$-order accuracy in time. The HLLC-ADC Riemann solver [69] is used to avoid numerical carbuncle phenomena, which can severely degrade the solution in strongly shocked reacting flows. The code implements block-based adaptive mesh refinement, which is critical for resolving the complex multi-scale structure of detonations. All calculations are performed in rectangular channels with isothermal, no-slip boundary conditions.

All simulations use modern multi-step chemical kinetics mechanisms. In particular, calculations, which consider hydrogen and light hydrocarbons, were performed using the FFCM-1 family of skeletal mechanisms [70], while simulations with heavy hydrocarbons (JP-10 and $n$-dodecane) used the reduced HyChem mechanisms [71, 72]. The physical model includes detailed multi-species transport description based on mixture averaging. Further details of the reacting-flow equations solved and the physical model implemented can be found in Refs. [73, 74].

Table 2 lists all reactive mixtures considered below along with their key properties, including upstream pressure, $P$, temperature, $T$, and equivalence ratio, $\phi$. For simulations in regular or moderately irregular mixtures, it also provides the computed average detonation cell width, $\lambda_c$, along with the corresponding minimum and maximum values, in the format $\lambda_{min}^{max}$. The procedure used to obtain these values of $\lambda_c$ is described in §3.

Table 2 also lists the characteristic experimental detonation cell size, $\lambda_{exp}$, for each mixture, which was used to set the appropriate channel size in the simulations. Presently, there is no well-defined and universally agreed upon standard for measuring and reporting detonation cell data, which introduces con-

Table 2: Reactive mixtures considered. $P$, $T$: upstream pressure and temperature; $\phi$: mixture equivalence ratio; $\lambda_c$: numerical cell size; $\lambda_{exp}$: experimental cell size based on the references listed in the last column.

| # | Mixture | P (bar) | T (K) | $\phi$ | $\lambda_c$ (cm) | $\lambda_{exp}$ (cm) | Reference |
|---|---|---|---|---|---|---|---|
| 1 | $2H_2/O_2/3.76N_2$ | 1.00 | 298 | 1.0 | $0.32^{0.39}_{0.11}$ | $1.21^{1.78}_{0.60}$ | [41–47] |
| 2 | $2H_2/O_2/3.76N_2$ | 0.25 | 293 | 1.0 | $2.2^{2.50}_{1.20}$ | $2.4 \pm 0.8$ | [46] |
| 3 | $2H_2/O_2/3.76N_2$ | 1.00 | 650 | 1.0 | $0.42^{0.60}_{0.35}$ | $0.45^{0.50}_{0.40}$ | [41, 42, 48] |
| 4 | $H_2/O_2/3.76N_2$ | 1.01 | 298 | 0.5 | — | $10.15^{12.69}_{7.61}$ | [49, 50] |
| 5 | $H_2/O_2/3.76N_2$ | 1.01 | 650 | 0.5 | $1.0^{1.1}_{0.80}$ | $1.92 \pm 0.48$ | [42] |
| 6 | $CH_4/2O_2/7.52N_2$ | 1.01 | 298 | 1.0 | — | $26.5^{32.50}_{17.50}$ | [51, 52] |
| 7 | $C_2H_4/3O_2/11.28N_2$ | 5.00 | 298 | 1.0 | — | $0.8^{\dagger}$ | [53] |
| 8 | $0.7C_2H_4/3O_2/11.28N_2$ | 1.00 | 298 | 0.7 | — | $8.0^{9.00}_{7.00}$ | [54–56] |
| 9 | $3C_2H_2/2.5O_2/9.4N_2$ | 1.01 | 293 | 3.0 | $0.2^{0.5}_{0.1}$ | 1.16 | [54] |
| 10 | $1.4C_{10}H_{16}/14O_2/52.64N_2$ | 1.00 | 353 | 1.4 | — | $7.41^{9.90}_{4.00}$ | [57] |
| 11 | $n C_{12}H_{26}/18.5O_2/69.56N_2$ | 1.01 | 420 | 1.0 | — | $4.0^{\ddagger}$ | [57, 58] |

$^{\dagger}$Cell size estimated via extrapolation from the cell size data reported in Ref. [53]. $^{\ddagger}$Cell size estimated using the linear correlation between the cell size and the ZND induction length for similar hydrocarbons in Refs. [57, 58].

siderable uncertainties. For instance, very often, experimental soot foils are not available, the precise method for determining the cell sizes is not described, and associated uncertainties of the reported values are not provided. In many studies, it is also not possible to determine the impact of wall losses on the reported cell sizes, for example, due to the lack of reliable detonation speed measurements. In particular, any detonation velocity deficit relative to $D_{CJ}$ would directly indicate the importance of wall losses. Lastly, whenever the detonation velocity is not provided, it is also not possible to determine whether the detonation has reached a steady state.

Mixtures listed in Table 2 were selected based on the availability of reliable experimental measurements of $\lambda_{exp}$ in the literature, with the exception of mixtures 7, for which $\lambda$ is based on the extrapolation of the existing measurements at lower pressures [53], and 11, for which previous phenomenological correlation for heavy hydrocarbons was used [57]. In particular, the following selection criteria were applied: (i) First, we selected measurements, for which a reliable detonation speed within a few percent of $D_{CJ}$ (with an associated uncertainty) was reported. (ii) We considered only measurements obtained in channels, in which the smallest dimension in the case of rectangular channels or the tube diameter was at least a factor of $3-4$ greater than the reported cell size. Both of these criteria provide some degree of confidence that the detonation in an experiment was relaxed and its structure was not influenced by the wall losses. These criteria, however, significantly restrict the list of experimental studies, which are suitable for a direct comparison with simulations.

If, for a particular mixture, multiple cell size measurements are available that satisfy these criteria, we list the average value, along with the minimum and maximum reported values, in the format $\lambda^{max}_{min}$. However, if only one measurement exists and the corresponding experimental uncertainty $\delta\lambda$ is provided, it is listed as $\lambda \pm \delta\lambda$. Detailed summary of the experimental studies listed in Table 2 is provided in Table S2 in the Supplementary Material.

Table 3: Properties of the reactive mixtures listed in Table 2: $D_{CJ}$, $\epsilon$, $\gamma_{VN}$, $\Delta_i$, and $\Delta_e$ are, respectively, the CJ detonation speed, the effective activation energy (eq. 1), the adiabatic index at the VN state, the ZND induction length, and the exothermic pulse width. Numbers in parentheses are powers of 10.

| # | $D_{CJ}$ (km/s) | $\epsilon$ | $\gamma_{VN}$ | $\Delta_i$ (cm) | $\Delta_e$ (cm) |
|---|---|---|---|---|---|
| 1 | 1.98 | 7.84 | 1.32 | 2.16 (-2) | 5.64 (-3) |
| 2 | 1.94 | 6.60 | 1.32 | 8.54 (-2) | 2.57 (-2) |
| 3 | 1.93 | 5.87 | 1.31 | 2.53 (-2) | 1.11 (-2) |
| 4 | 1.62 | 35.97 | 1.33 | 3.20 (-1) | 1.48 (-2) |
| 5 | 1.63 | 6.72 | 1.31 | 5.32 (-2) | 1.81 (-2) |
| 6 | 1.81 | 15.17 | 1.26 | 2.51 (+0) | 2.74 (-2) |
| 7 | 1.86 | 13.06 | 1.25 | 2.53 (-2) | 5.16 (-3) |
| 8 | 1.70 | 15.64 | 1.27 | 1.94 (-1) | 2.24 (-2) |
| 9 | 1.99 | 4.21 | 1.23 | 1.94 (-2) | 1.62 (-2) |
| 10 | 1.83 | 10.83 | 1.23 | 2.43 (-1) | 3.00 (-2) |
| 11 | 1.79 | 9.76 | 1.23 | 1.70 (-1) | 2.49 (-2) |

Finally, Table 3 lists key properties of the CJ detonations in each of the mixtures given in Table 2, including the CJ detonation speed, $D_{CJ}$, effective activation energy, $\epsilon$ (see eq. (1) below), the adiabatic index at the von Neumann (VN) state, $\gamma_{VN}$, as well as the ZND induction length, $\Delta_i$, and exothermic pulse width, $\Delta_e$.

## 3. Structure and Dynamics of Planar Multi-Dimensional Detonations

The defining characteristic of a realistic multi-dimensional detonation is the intrinsic instability of its front [75], which is illustrated in Fig. 1a showing the volume rendering of the leading shock in a stoichiometric $C_2H_4$/air at 5 bar (mixture 7, Table 2). Even in this idealized, freely propagating detonation in a rectangular channel, the front is not planar but has a characteristic cellular structure formed by the complex pattern of regions with weaker (underdriven) and stronger (overdriven) shocks. The Mach numbers of the different sections of the front can vary in the range $\sim 4-9$ for typical fuels, creating pressures

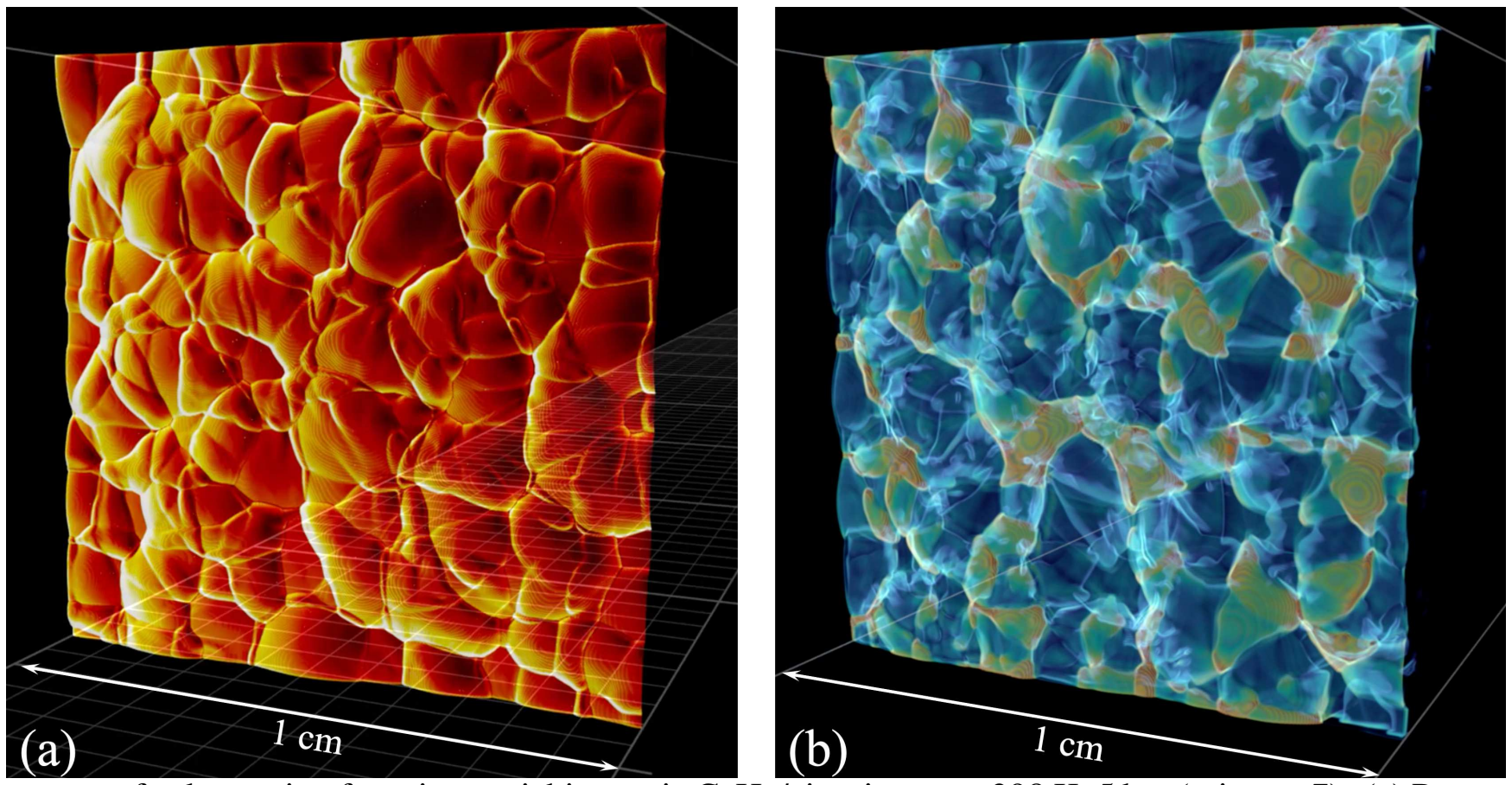


Fig. 1: 3D structure of a detonation front in a stoichiometric $C_2H_4$/air mixture at 298 K, 5 bar (mixture 7). (a) Pressure isosurface of 10 bar shown with varying opacity using a ray-tracing visualization technique. (b) Temperature gradient magnitude. Shown is the semitransparent detonation volume with varying opacity using a ray-tracing visualization technique. Colors correspond to the gradient magnitude increasing from blue to red.

that can reach thousands bar (even at moderate upstream pressures of a few bar) and temperatures in excess of 4000 K. The internal structure of a detonation (Fig. 1b) is composed of a complex web of transverse waves that move inside the front. Periodic collisions of such transverse waves create regions of large temperature gradients associated with the bulging overdriven sections of the front. Even with a relatively modest spatial resolution of $\Delta_i/\Delta x = 24.5$ and $\Delta_e/\Delta x = 5$, this simulation consisted of $\sim$ 3 billion cells, it was performed on 65,000 cores and required $\approx$ 25 million CPU hours, which illustrates the typical cost of modern high-fidelity detonation simulations in realistic fuels.

The instability of the front is the main mechanism that determines the key dynamical properties of a multi-dimensional detonation, namely its ability to propagate in narrow channels [44], survive its emergence into larger unconfined spaces [76], or propagate over obstacles [77]. It is impossible to understand the nature of a detonation, and to be able to control and manipulate it in practical settings, without understanding the physics of this instability.

A convenient way to visualize such a complex detonation structure is through the so-called soot foils. It was first observed in early detonation experiments that a detonation propagating in a channel with walls covered with soot leaves a characteristic cellular pattern on the channel walls [78]. The exact mechanism that forms the pattern seen in the soot foils has been investigated in several studies [79–81], yet it remains not fully understood. Different quantities, including pressure, enstrophy, heat release rate, and shear stress, measured on the channel walls have been used in the past as the numerical soot foil surrogates [82–86]. While they can produce somewhat different morphology of the detonation cells, crucially they all give the same cell size. In numerical simulations, synthetic soot foils are typically obtained by recording the maximum pressure at each point on the channel wall, which results in patterns morphologically most similar to those observed in the experiments.

A soot foil allows the characteristic detonation scales to be identified easily, in particular characteristic spacing between the transverse waves in the front, or equivalently the frequency of their collisions. Thus, it simultaneously captures both temporal evolution of the front and its spatial structure. Important dynamical aspects of the front, such as detonation failure/reignition, can also be recorded in the soot foil traces. Other imaging techniques exist and are often used in experiments [15, 87], for example, based on streak camera imaging [87], although such techniques are typically more complex.

An example of a numerical soot foil produced in a stoichiometric $H_2$/air mixture at atmospheric pressure and an elevated temperature of 650 K is shown in Fig. 2a,b. Both 2D and 3D calculations were performed in a rectangular channel 1.6 cm wide and 0.4 cm deep in 3D. In 2D calculations (Fig. 2a), the unstable front creates a characteristic regular pattern of rhomboidal cells, which has been seen in numerous previous studies [16, 88–90]. Mixtures producing such regular patterns are often characterized as regular, or 'stable'. However, the overall morphology of cells can differ considerably in a more realistic 3D case (Fig. 2b), with individual cells having a much more irregular and complex structure, although the overall rhomboidal shape can still be identified.

In order to characterize the cell size in the synthetic soot foils quantitatively, the following approach was used. First, for a given soot foil, we considered the last section, the length of which was three times the domain width, $L$. This section was split into three sub-sections, each one with length $L$. An open-source, computer-vision-based processing tool developed by Jalontzki et al. [91] was applied to analyze each of these three sub-sections. This processing tool creates a histogram of cell sizes, providing the average value along with the corresponding range. We ensured that

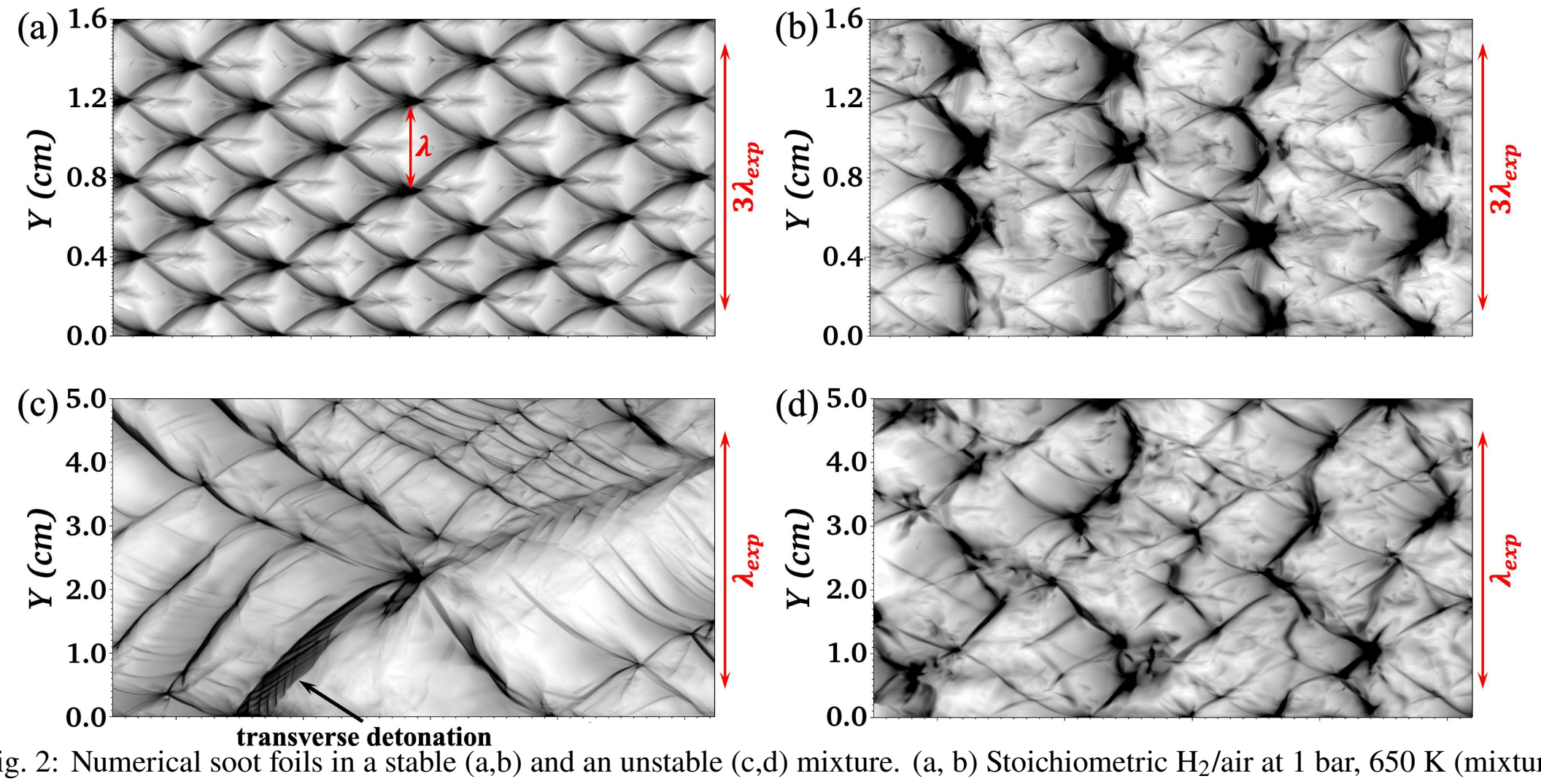


Fig. 2: Numerical soot foils in a stable (a,b) and an unstable (c,d) mixture. (a, b) Stoichiometric $H_2$/air at 1 bar, 650 K (mixture 3). Shown is the logarithm of maximum pressure recorded at each point. Domain width is equal to $\approx 3.6\lambda_{exp}$ [42]. (c, d) Stoichiometric *n*-dodecane/air at 1 bar, 420 K (mixture 11). Image height is equal to 1.25 times the cell size predicted based on prior phenomenological correlations [57, 58]. In all panels, detonation propagates from right to left.

the average value remained constant in each of these three sub-sections. This demonstrated temporal convergence of the cellular structure, which, along with the requirement for the detonation speed to be within $<1\%$ of $D_{CJ}$, showed that the front was fully relaxed.

Both in the 2D and 3D simulations in Fig. 2a,b, four cells are formed in the 1.6 cm wide channel, with the average value of $\lambda_c = 0.42$ cm (2D case) in close agreement with the experimental value $\lambda_{exp} = 0.45$ cm (Table 2). This particular preheated case is the only $H_2$/air mixture at atmospheric pressure known to authors, in which numerical simulations predict correct detonation cell size. Similar agreement (Table 1) was previously observed by Smirnov et al. [20] using several multi-step chemical mechanisms different from the FFCM-1 used here. Also, Fig. 2b (and Fig. 3c,d below) shows the first 3D simulation of this mixture, which is described in further detail by Ghosh et al. [40]. Virtually all other prior numerical simulations, in which a detonation cell size was in agreement with the experimental values, were performed at pressures significantly below atmospheric [16, 88] (also see Table 1 and Fig. 7b below).[1]

Note that both in the 2D and 3D simulations, the channel width is the same to allow direct comparison between them, and it is slightly less than $4\lambda_{exp}$. This choice of the domain width was primarily dictated by the large computational cost of the 3D calculation, which involved 18 billion grid cells. To ensure that the observed cellular structure was not a result of mode locking, we performed an additional 2D calculation with a three times larger width, namely 4.8 cm instead of 1.6 cm. Corresponding soot foil is shown in Fig. S2a in the Supplementary Material. For comparison, Fig. S2b also shows the same soot foil as in Fig. 2a on the same scale as Fig. S2a. Note that the soot foil in Fig. S2a is not only wider, but it also shows longer time evolution. In the wider channel, $\lambda_c = 0.42^{0.59}_{0.33}$ cm compared to $\lambda_c = 0.42^{0.60}_{0.35}$ cm (see Table 1), i.e., the average value was the same with the somewhat smaller minimum and maximum values, thus demonstrating that the cell size in the original calculation was not mode locked.

Close agreement of the characteristic numerical detonation scales shown in Fig. 2a,b with the experimental values suggests that the internal front structure in these calculations does reflect the reality. Figure 3 shows the corresponding structure of the front in the 2D (panels a,b) and 3D (panels c,d) simulations.

The 2D front has a regular structure, with four very similar regions, which create four detonation cells. One of such regions is shown as a close-up view in panel (b). The main underlying cause of the front instability is the high temperature sensitivity of the reactions, which is typical for almost all realistic fuels [16, 75]. If the shock strength decreases locally, resulting significant increase in the time for the mixture to ignite creates a large region of hot, compressed reactants. Two such fuel pockets formed behind weak, underdriven sections of the front can be seen at the top and bottom of Fig. 3b. Two transverse waves, consist-

[1] Figure S1b in the Supplementary Material shows an additional case performed at the temperature of 373 K intermediate between 298 K (mixture 1; see Fig. 7a below) and 650 K (mixture 3). Figure S1b shows progression in cell structure with increasing upstream temperature for all three mixtures demonstrating that even a modest increment in temperature from 298 K to 373 K increases both cell regularity and cell size relative to the experimental value, compared to mixture 1, even though cells are still smaller than $\lambda_{exp}$. Also note that the cell size in Fig. S1b is similar to the one reported in Ref. [20] at 400 K, namely 0.4 cm (Table S1).

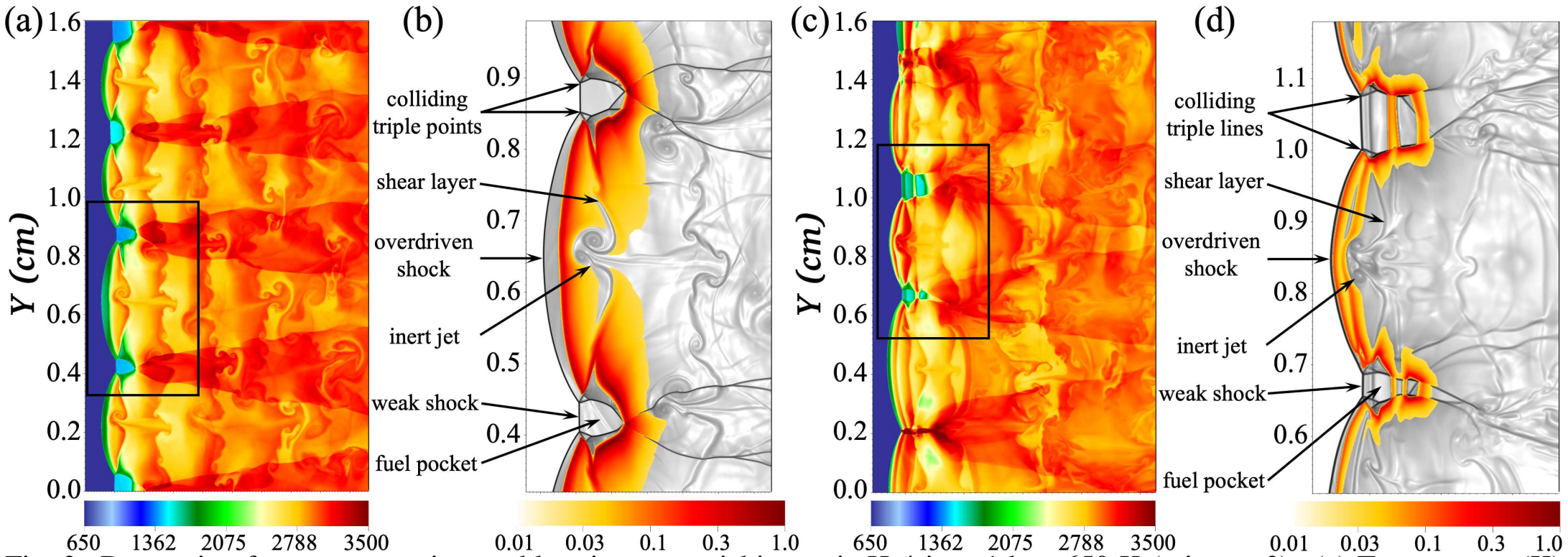


Fig. 3: Detonation front structure in a stable mixture: stoichiometric $H_2$/air at 1 bar, 650 K (mixture 3). (a) Temperature (K) in the 2D calculation (cf. Fig. 2a). (b) Synthetic schlieren ($\log|\nabla\rho|$, grayscale) and normalized heat release rate (color) in the region marked with a black box in panel (a). (c) Temperature distribution (K) in the midplane of the domain in a 3D calculation (cf. Fig. 2b). (d) Synthetic schlieren ($\log|\nabla\rho|$, grayscale) and normalized heat release rate (color) in the region marked with a black box in panel (c). Heat release rate is normalized by (b) $1.1\times10^{17}$ erg/g s and (d) $5.1\times10^{17}$ erg/g s. Domain is 1.6 cm, or $3.6\lambda_{exp}$ (Table 2), wide in both 2D and 3D, and 0.4 cm, or $\approx0.9\lambda_{exp}$, deep in 3D.

ing of the triple points and attached transverse shocks, move toward each other in each of these pockets, rapidly consuming unburnt compressed fuel. Eventual collision of these transverse waves forms an overdriven section of the front [16, 75], which accelerates upstream and thus pushes the detonation front forward. This overdriven shock develops a convex shape and, thus, it forms the next generation of transverse waves, which spread laterally into the neighboring underdriven sections of the front. Over time, the overdriven shock weakens to the point when it no longer ignites the fuel sufficiently fast, thus creating a new pocket of fuel, through which transverse waves can propagate, collide, and form the next overdriven shock. This creates a natural life cycle of a detonation, which supports its stable propagation [75]. Such cyclic dynamics is characteristic of stable mixtures, which produce regular cellular patterns (Fig. 2a,b).

Figure 3c,d shows the corresponding detonation structure in the midplane of a 3D simulation. Qualitatively, all the same key features are present in 3D as in 2D, including fuel pockets, colliding transverse waves, and the convex overdriven shock section formed as a result of such collision. At the same time, the 2D geometry tends to oversimplify the overall structure by regularizing main dynamical features. For example, Fig. 3b shows a prominent remnant of a jet, which was initially formed by the converging flows behind the two transverse waves. The formation mechanism of such jets and their importance for the generation and support of the overdriven shock sections are not fully understood [31, 92, 93]. While in 2D such jets are well defined with a clear structure, in a more realistic 3D geometry, they are less pronounced, though they may still be identified in regular mixtures. This is due to the fact that in 3D, unlike 2D, transverse wave collisions do not occur at a point at discrete time instants. Instead of triple points, triple lines move constantly out of plane interacting along complex multi-dimensional trajectories. As a result, even though jet-like features may form through flow collisions, they often have irregular structure, and they continuously change their orientation as they follow the motion of the colliding flows behind the 3D transverse waves. The presence of such out-of-plane motion of the transverse waves manifests itself as the characteristic parallel shocks, as well as parallel structures of high heat release rate, which can be seen in Fig. 3d. Such multiple streak-like features, which closely follow each other, are never observed in the 2D geometry.

Figure 2c,d also gives an example of a very different irregular detonation type, which is often associated with 'unstable' mixtures. It shows the numerical soot foils in a stoichiometric *n*-dodecane/air at 1 bar, 420 K (mixture 11). The 2D calculation (Fig. 2c) was performed in a 10 cm wide channel, while the 3D channel (Fig. 2d) had a cross-section of $5\times1$ cm.

The 2D cellular structure is highly irregular with a widely varying cell size. The overall front dynamics also is much more unstable with a pronounced detonation quasi-failure, which can be seen as the extended light region. During such quasi-failure, cells effectively disappear as a result of the transverse-wave failure in a section of the front. This is followed by the formation of a transverse detonation, which sweeps across the front re-energizing it. Such highly unstable dynamics, including the formation of transverse detonations, was observed in prior 2D simulations [94].

Similar highly irregular structure is characteristic not only of heavy hydrocarbon fuels but also light hydrocarbons and, in general, any reactive mixture characterized by high activation energy. As an example, Fig. 4 shows the detailed front structure in the lean ($\phi = 0.7$) $C_2H_4$/air at 1 bar, 298 K (mixture 8). In this case, detonation has a much more complex structure than in a regular mixture (Fig. 3). In 2D (Fig. 4b), transverse waves can take many forms, including transverse detonations, strong transverse oblique reacting shocks, transverse 'fingers', and turbulent deflagration-like waves propagating into the

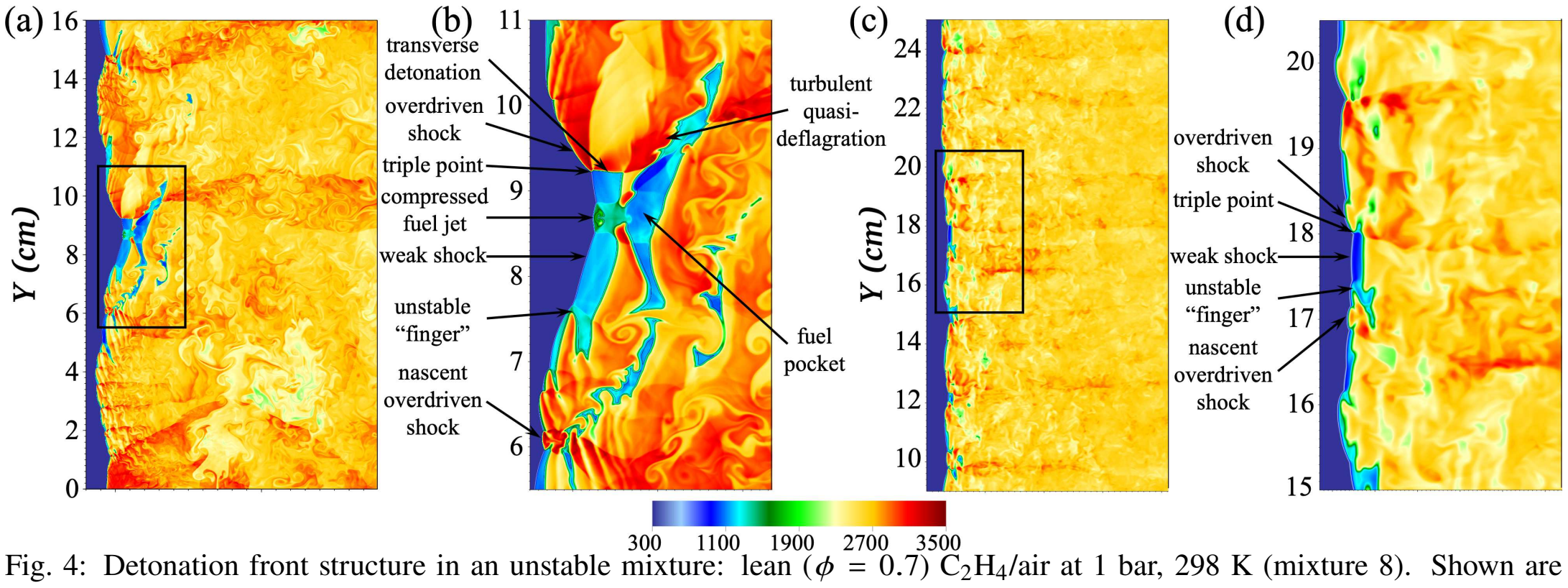


Fig. 4: Detonation front structure in an unstable mixture: lean ($\phi$ = 0.7) $C_2H_4$/air at 1 bar, 298 K (mixture 8). Shown are the temperature distributions in the 2D (a, b) and the midplane of the domain in the 3D (c, d) calculations. Panels (b) and (d) correspond to the regions marked with a black box in panels (a) and (c), respectively. Domain is 30.48 cm, or $\approx 3.8\lambda_{exp}$ (Table 2), wide in both 2D and 3D, and 2.54 cm, or $\approx 1/3\lambda_{exp}$, deep in 3D.

regions of compressed reactants formed behind the underdriven shock sections. Transverse detonations (also seen in Fig. 2c) can form when an extended fuel pocket is created as a result of a local quasi-failure of the front seen in Fig. 4b [95, 96]. Otherwise, instead of a pair of triple points formed by an overdriven section, as typically seen in stable mixtures, multiple unstable finger-like structures can move in the same direction. These are formed by the individual triple points, which trigger burning as they propagate, though they often fail, with reactions gradually becoming decoupled. Finally, consumption of large fuel pockets can also in part be driven by turbulent quasi-deflagration-like waves formed in high-speed shear layers deeper inside the detonation [76, 83, 85, 94, 96, 97, 97, 98]. The role of such quasi-deflagrations is not well understood [76, 85, 97, 97], and it is possible that such structures may be exaggerated by the 2D geometry. All these flow structures span a huge range of spatial and temporal scales from the macro-scales effectively down to the continuum limit, which creates a unique set of challenges both for modeling and experimental studies.

In contrast, 3D simulations in irregular mixtures generally exhibit much less unstable behavior. Figure 2d shows the soot foil in the 3D simulation of the same $n$-dodecane/air mixture shown in Fig. 2c. While the cell shapes are quite irregular, their sizes vary much less than in 2D. Furthermore, quasi-failures/re-ignitions and formation of transverse detonations are generally not observed in 3D simulations.

Such more stable dynamics is illustrated in Fig. 4c,d, which shows the front structure in the midplane of a 3D detonation in the same $C_2H_4$ mixture as in Fig. 4a,b. In this calculation, the 3D domain cross-section is based on the detonation tube facility at the US Air Force Research Laboratory [99, 100]. The domain is quite narrow with the depth-to-width ratio of 1/12 with the narrow dimension equal to $\approx 1/3\lambda_{exp}$. This would suggest a quasi-2D behavior. Even in such a narrow 3D channel, many of the features seen in 2D, including transverse detonations and multiple unstable fingers, are not present, even though the front is still quite irregular.

In general, 2D simulations in unstable mixtures are much less representative of the actual detonation dynamics than in stable ones. They often exaggerate the unstable behavior producing quasi-failures and re-ignitions, which are typically not seen in 3D and which can trigger unphysical global detonation failure. However, crucially, the characteristic average cell size is generally comparable between 2D and 3D in both stable and unstable mixtures.

Most importantly, Figs. 2c,d and 4 demonstrate the main problem. With a few exceptions, current multi-dimensional simulations, which use the state-of-the-art thermochemical models, such as the ones employed here, are generally unable to predict the detonation cell size accurately for most practical fuels under practically relevant conditions. In particular, cell size for the $n$-dodecane/air mixture (Figs. 2c,d) is expected to be $\lambda \approx 4$ cm based on the phenomenological correlation for hydrocarbons [58], i.e., there should be approximately one cell in a 5 cm wide channel shown here. Actual cell size, however is considerably smaller, even in 3D. Similarly, in the lean $C_2H_4$/air mixture, experimental cell size $\lambda_{exp} \approx 8$ cm [54], and thus a 16 cm section of the front shown in Fig. 4 should contain approximately two cells, and thus two pairs of transverse waves. Instead, not only in 2D, but even in 3D, the characteristic spacing between transverse waves is much smaller. In particular, Fig. 4d shows that such characteristic separation is less than the domain depth equal to 2.54 cm. As a result, there are multiple transverse waves in the narrow transverse direction, which allows the detonation front in the simulation to develop a fully 3D structure even in such a narrow channel.

In addition to $n$-dodecane and ethylene, a few more examples of the detonation structure in other hydrocarbon fuels, from light to heavy, are given in Fig. 5. Shown are the soot foils in the 2D simulations in a stoichiometric $CH_4$/air mixture at 298 K (mixture 6; Fig. 5a), rich ($\phi$ = 3) $C_2H_2$/air mixture at 293 K (mixture 9; Fig. 5b), and rich ($\phi$ = 1.4) JP-10/$O_2$/$N_2$

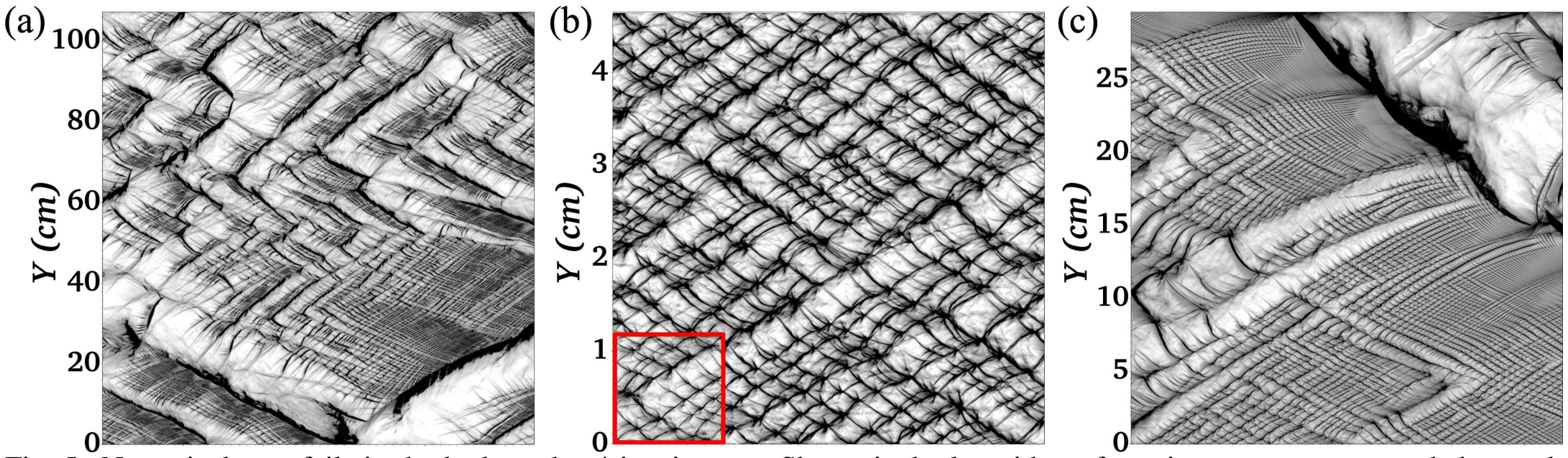

Fig. 5: Numerical soot foils in the hydrocarbon/air mixtures. Shown is the logarithm of maximum pressure recorded at each point. (a) Stoichiometric $CH_4$/air at 1 bar, 298 K (mixture 6). (b) Rich ($\phi$ = 3) $C_2H_2$/air at 1 bar, 293 K (mixture 9). (c) Rich ($\phi$ = 1.4) JP-10/$O_2$/$N_2$ at 1 bar, 353 K (mixture 10). Red insert in panel (b) shows the soot foil from a simulation with four times higher resolution $\Delta x$ = 700 nm. In all panels, domain width is $4\lambda_{exp}$ (Table 2) and detonation propagates from right to left.

mixture at 353 K (mixture 10; Fig. 5c). All cases were carried out at atmospheric pressure. Again, domain width was four times the published experimental cell width for each mixture [52, 54, 57] (Table 2).

Similar to *n*-dodecane and $C_2H_4$, Fig. 5 shows that cell sizes in these three mixtures are significantly smaller than the experimental values, and cells in some regions of the $CH_4$ and JP-10 soot foils are virtually indistinguishable. More specifically, in the $C_2H_2$/air mixture (panel b), computed cell size is $\lambda_c = 0.2^{0.5}_{0.1}$ cm, or almost a factor of six smaller than $\lambda_{exp}$ (Table 2). In $CH_4$ and JP-10 (panels a, c), the detonation dynamics is extremely unstable, with cell sizes varying over a wide range and with transverse detonations present (dark angled regions; cf. Fig. 2c). As discussed above, such highly unstable dynamics is likely not representative of a 3D detonation. At the same time, based on the *n*-dodecane and $C_2H_4$ cases discussed above, cell sizes in 3D can be expected to be comparable to those shown in Fig. 5.

It is important to note that the quantitative analysis of the detonation cell sizes, which is based on various automated approaches, such as the one developed by Jalontzki et al. [91] and used here, is generally applicable only to regular or moderately irregular cellular structures, e.g, shown in Fig. 2a (mixture 3) or Fig. 5b (mixture 9). In such mixtures, one can define individual cells and, therefore, meaningfully extract their average size (e.g., see mixtures 1-3, 5, and 9 in Table 2). In contrast, for extremely irregular mixtures, such as in Figs. 2c,d and 5a,c (as well as mixture 4 discussed below), it is virtually impossible to define a cell unambiguously due to the presence of large regions of detonation failure, as well as the lack of clearly defined cell boundaries. As a result, in such cases, automated tools designed to extract cell sizes generally fail. At the same time, all cell-like structures in the soot foils in panels (a) and (c) are much smaller than 1/4 of the channel width based on the experimentally measured values of $\lambda_{exp}$.

Overall, there are far fewer detonation simulations of hydrocarbons in air compared to hydrogen mixtures,[2] and similar to hydrogen, most existing studies are performed at pressures significantly below atmospheric [14, 29, 30, 101]. However, even under such low-pressure conditions, the numerical cell sizes are often smaller than $\lambda_{exp}$ [14, 29, 102]. At atmospheric pressure, Liu & Zhang [32] reported cells in a JP-10/air mixture close to the experimental values. However, the detonation structure appears to evolve with time, suggesting that a steady state has not been reached. Furthermore, the size of the detonation cells decreases with increasing resolution. This decrease in $\lambda$ with resolution in unstable mixtures was first shown by Taylor et al. [18]. Therefore, robust conclusions regarding agreement with the experimental results cannot be made, unless the numerical solution is shown to be converged both spatially and temporally. In summary, in contrast to hydrogen, there is currently no conclusive evidence of the ability of the multi-dimensional numerical simulations of detonations in hydrocarbon fuels in air, which use realistic multi-step chemical kinetics models, to produce the correct detonation cell size under the atmospheric (or higher-pressure) conditions.

## 4. Possible Sources of the Discrepancy between Simulations and Experiments

What can be the source of such stark disagreement between simulations and experiments? First, we emphasize that one cannot rule out the possibility that the source may in fact be the numerical approach. In particular, 2D vs. 3D geometry (Fig. 2c,d), varying channel length (up to $\sim 160\lambda_{exp}$) or width (beyond $4\lambda_{exp}$; see Fig. S2 in Supplementary Material), or insufficient numerical resolution cannot explain such dramatically small cells, as seen in Fig. 5 [18, 20]. Significant wall losses (thermal and viscous) can increase the detonation cell size in the near-limit regimes of detonation propagation with significant velocity deficits of $\gtrsim 10-14\%$ [83, 98]. For this reason, as discussed in §2, we specifically perform comparison only between numerical and experimental studies, in which evidence exists of the minimal impact of wall losses on the measured detonation structure. Furthermore,

[2] Note that study by Dushe et al. [31] (Table 1) considered $O_2$-enriched mixtures of $CH_4$ and $C_2H_2$.

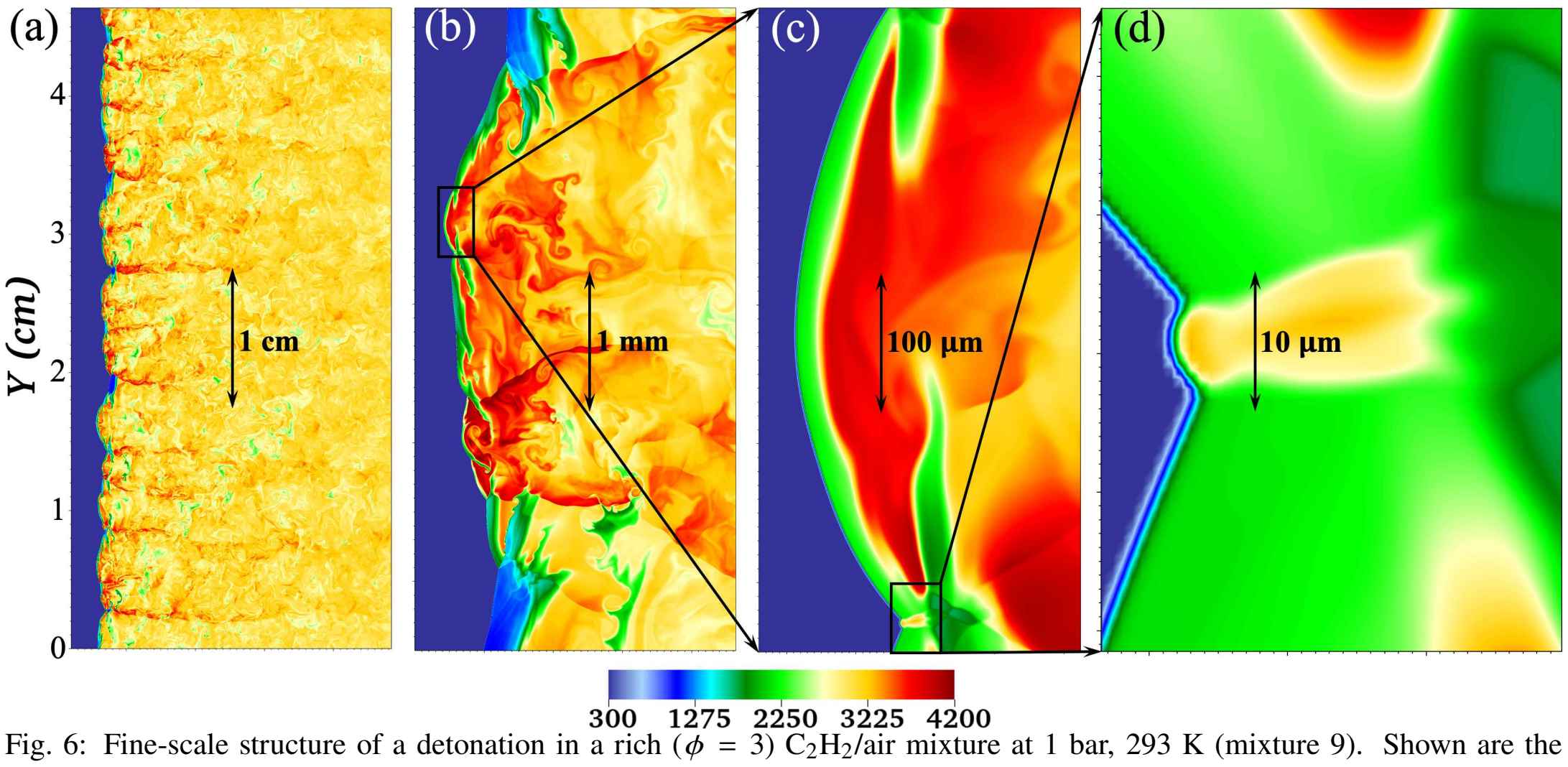


Fig. 6: Fine-scale structure of a detonation in a rich ($\phi$ = 3) $C_2H_2$/air mixture at 1 bar, 293 K (mixture 9). Shown are the temperature distributions (K) in regions 4.64 cm (a), 4.64 mm (b), 464 μm (c), and 46.4 μm (d) in height. Panel (a) is from a lower resolution simulation with $\Delta x$ = 2.8 μm (cf. Fig. 5b). Panels (b)-(d) are from a higher-resolution simulation with $\Delta x$ = 700 nm (cf. red insert in Fig. 5b). Panels (c) and (d) correspond to the regions marked with black rectangles in panels (b) and (c), respectively.

in all cases considered here, effects of wall losses were also minimal with the detonation velocity being within $\lesssim 1\%$ of $D_{CJ}$.

Effect of resolution on the detonation structure is illustrated in Fig. 5b for a rich $C_2H_2$/air mixture.The inset in Fig. 5b marked with a red square shows the soot foil from a simulation with a four times higher resolution, $\Delta x$ = 700 nm instead of 2.8 μm. Due to the high computational cost, this simulation was performed in a four times narrower domain. Such small computational grid size is approximately ten times the mean free path in the upstream gas, which is close to the continuum limit, and any further increase in resolution in a simulation based on the Navier-Stokes equations would be unphysical. Qualitatively, Fig. 5b shows that even at such high resolution, the detonation cell size does not increase, but instead cells become slightly smaller as finer features appear. More quantitatively, the computed cell size in the coarse-grid calculation was $\lambda_c = 0.2_{0.1}^{0.5}$ cm (Table 2), while in the highly-resolved case $\lambda_c = 0.2_{0.06}^{0.45}$ cm. Thus, the average value did not change with increasing resolution, though the minimum observed cell size became somewhat smaller as a result of finer structures being formed at higher resolution. Such formation of progressively finer structures with increasing resolution in moderately irregular mixtures was previously shown by Taylor et al. [18] for mixture 1, i.e., the canonical stoichiometric, atmospheric $H_2$/air (also see Fig. 7a below). Finally, we note that for the preheated stoichiometric mixture 3, i.e., regular mixture with the correct cell size, numerical convergence in the 2D and 3D calculations was discussed in detail by Ghosh et al. [40] (also see Ref. [103].

Part of the reason why it is hard to achieve grid convergence in such unstable detonations is that flow structures form on all scales in a simulation, down to the computational grid scale. Such smaller structures contribute to the instability of the front, allowing progressively more transverse waves to form, and thus decreasing, rather than increasing the cell size. This is illustrated in Fig. 6, which shows the fine-scale structure of a detonation in the same rich $C_2H_2$/air mixture as in Fig. 5b. Panel (a) shows the full front in a lower-resolution calculation ($\Delta x$ = 2.8 μm) in a 4.64 cm channel. The front is highly irregular, which is typical for an unstable mixture. Panel (b) shows a region that is ten times smaller, namely 4.64 mm in height, in the higher resolution calculation with $\Delta x$ = 700 nm. Panel (c) shows a convex, overdriven section of the front $\approx$ 0.5 mm wide. The transverse wave pushed downward by this overdriven shock collided with another transverse wave near the bottom of panel (c), forming a nascent hot spot, which will evolve into a new overdriven section. Note that this high-pressure collision region shown in more detail in panel (d) is only $\approx$ 10 μm wide with a clearly defined structure. This shows that dynamically important structures can form on all scales allowed by the grid resolution down to $\sim$ 10 computational cells. Such an extremely wide range of physical scales (note that detonation in Fig. 6 spans four orders of magnitude in spatial scales), coupled with a complex, highly non-linear physical model, make a detonation, arguably, one of the most challenging problems in modern computational fluid dynamics. Finally, note that in contrast to unstable mixtures, convergence can be achieved in detonations with a highly regular structure, especially in mixtures, in which close agreement with experiments is observed [40]. It is not clear, however, what prevents progressively smaller structures from forming in such regular mixtures.

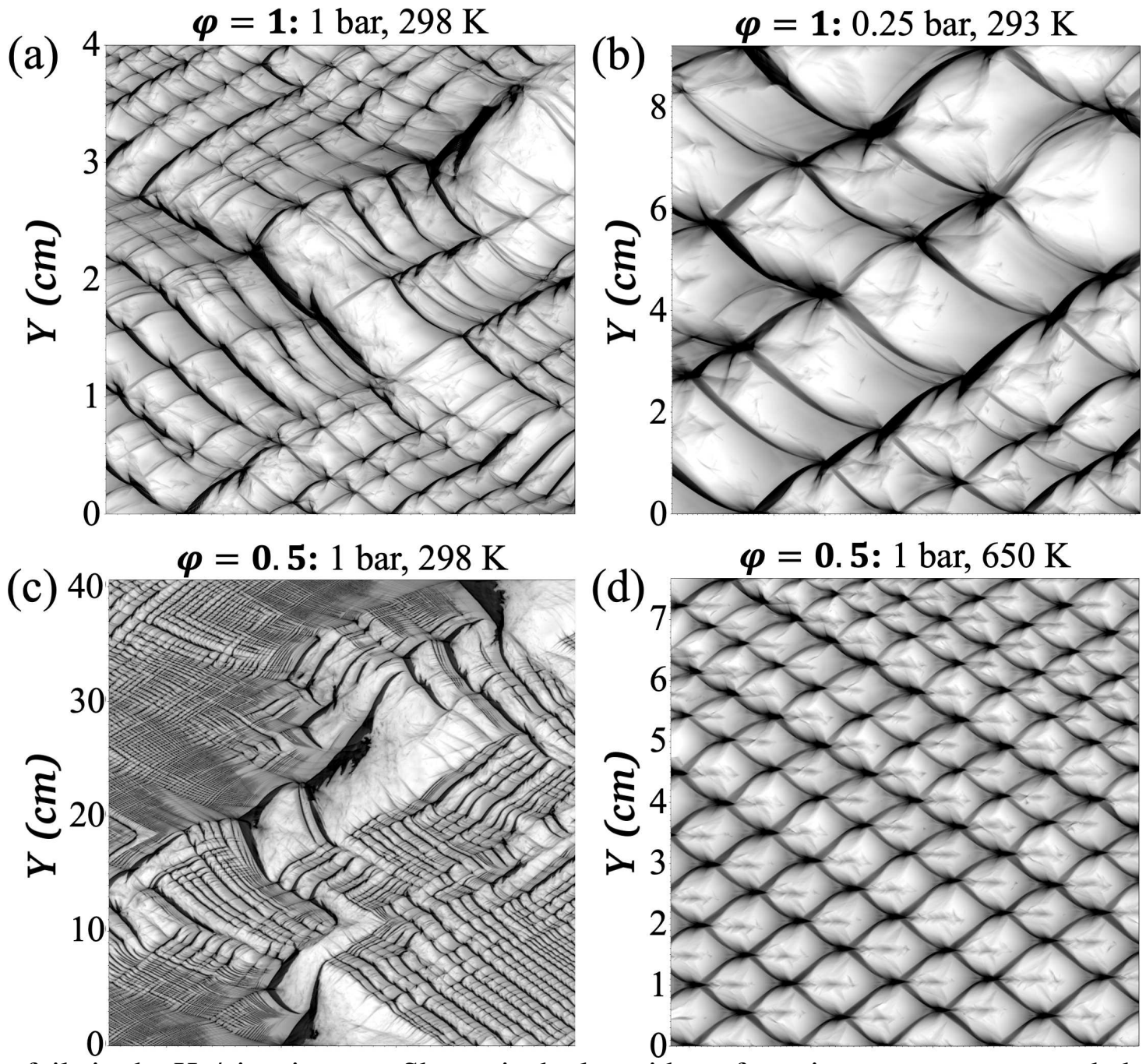


Fig. 7: Synthetic soot foils in the $H_2$/air mixtures. Shown is the logarithm of maximum pressure recorded at each point. $\phi = 1$: (a) 1 bar, 298 K (mixture 1); (b) 0.25 bar, 293 K (mixture 2). $\phi = 0.5$: (c) 1 bar, 298 K (mixture 4); (d) 1 bar, 650 K (mixture 5). In panel (a), domain width is $\approx 3.3\lambda_{exp}$, while in panels (b)-(d) it is $\approx 4\lambda_{exp}$ (Table 2). Detonation propagates from right to left.

The fact that a viscous Navier-Stokes model admits the formation of features approaching the continuum limit suggests two possibilities. The first possibility is that Navier-Stokes equations are not suitable for the description of the detonation dynamics if such dynamics can be controlled by scales close to or below the continuum limit. Alternatively, Navier-Stokes model is broadly valid. However, some (or all) components of the physical model are incorrect, and in reality a physical mechanism exists which stabilizes the front on such small scales, preventing fine structures from forming. Meaningful progress in our understanding of the detonation physics, and more broadly of high-speed combustion, requires systematic efforts to investigate these two possibilities. In recent years, attempts were made to go beyond the Navier-Stokes framework in detonation studies, for example, by employing discrete Boltzmann [107] and Direct Simulation Monte Carlo [108] methods. Here, we do not discuss this direction further, and instead we ask next how well the physical model, and more specifically the reaction kinetics, is understood and, ultimately, constrained.

## 5. Thermochemical State Space of the $H_2$/air Detonations

We start with detonations in hydrogen mixtures, in particular $H_2$/air. Hydrogen chemistry is simpler than that of hydrocarbons, it has been studied in great detail over the years [70, 72, 109–113], and thus it is instructive to consider potential causes of the disagreement between experiments and simulations in hydrogen mixtures first.

Figure 7a shows the soot foil in the most canonical case of a detonation in a stoichiometric $H_2$/air mixture at atmospheric conditions (mixture 1, Table 2). The average numerical cell size in this case is $\lambda_c = 0.32^{0.39}_{0.11}$ cm $\approx 1/4\lambda_{exp}$. This value closely agrees with the one previously obtained by Taylor et al. [18], and it is somewhat larger than the one reported by Smirnov et al. [20] (Table 1). Since this mixture is moderately irregular, the cell size varies considerably. We showed above that if the upstream mixture is preheated to 650 K, then $\lambda_c \approx \lambda_{exp}$ (Fig. 2a,b). Furthermore, if the pressure is reduced below 1 bar at the same upstream temperature of $\approx 300$ K, then close agreement with the experiments is also achieved. In particular, in a stoichiometric $H_2$/air mixture at 0.25 bar (mixture 2), $\lambda_c = 2.2^{2.5}_{1.2}$ cm, which is within 8% of $\lambda_{exp} \approx 2.4$ cm (Table 2; Fig. 7b). Similar agreement with experiments has been reported in several other previous studies of detonations in $H_2$/air at low-pressure conditions [16, 18, 88, 114]. Note also that in the preheated and low-pressure cases the cellular structure is quite regular.

Changing composition has a much more pronounced effect. Figure 7c shows the soot foil at atmospheric conditions, but in a lean ($\phi = 0.5$), rather than stoichiometric, mixture. The disagreement with the experiments becomes significantly worse. Not only does the dynamics become extremely unstable, but also the cell size varies over a wide range with large regions of the soot foil containing cells, which

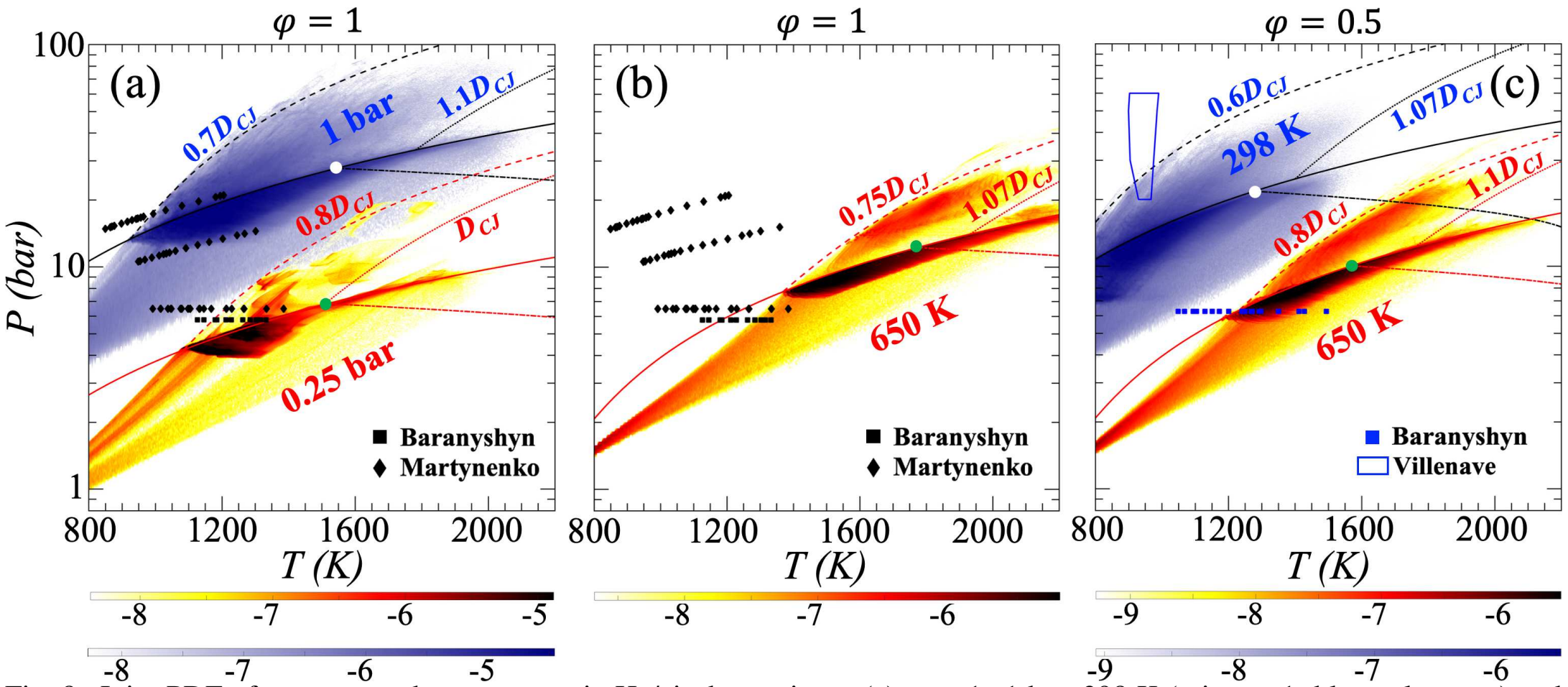


Fig. 8: Joint PDF of pressure and temperature in $H_2$/air detonations. (a) $\phi$ = 1: 1 bar, 298 K (mixture 1; blue color map) and 0.25 bar, 293 K (mixture 2; yellow-red color map). (b) $\phi$ = 1: 1 bar, 650 K (mixture 3). (c) $\phi$ = 0.5, 1 bar: 298 K (mixture 4; blue color map) and 650 K (mixture 5, yellow-red color map). In all panels, solid lines: primary shock adiabat; dashed and dotted lines: secondary adiabats corresponding to the lower and higher primary shock speeds, respectively, indicated in the figure; dash-dot lines: ZND solutions. White and green points: von Neumann state in a CJ detonation. Black and blue symbols mark conditions, at which experimentally measured $\tau_{ign}$ are available (cf. Figs. 9 and 10). Panels (a, b): black squares: Ref. [104], black diamonds: Ref. [105]. Panel (c): blue squares: Ref. [104], blue region: Ref. [106].

are almost indistinguishable. Similar to the $CH_4$, JP-10, and $n$-dodecane mixtures discussed above, it is not possible to define a meaningful average cell size in such a highly unstable mixture. However, again, all structures in the soot foil, aside from the extended regions of quasi-failure of the front, are much smaller than $\lambda_{exp} \approx 10$ cm, or $\approx 1/4$ of the channel width (Table 2). At the same time, similar to the $\phi = 1$ case, increasing the upstream temperature to 650 K at $\phi = 0.5$ brings the cell size to $\lambda_c = 1.0^{1.1}_{0.8}$ cm or within a factor of two of $\lambda_{exp} = 1.92$ cm (Table 2), which can be viewed as reasonable agreement given typical experimental uncertainties.

These results show that the disagreement between simulations and experiments exists under atmospheric conditions, and this disagreement becomes more severe when the composition differs from the stoichiometric one. This discrepancy disappears if the pressure is reduced or the upstream temperature is increased.

To better understand the differences between these cases, consider the thermodynamic state space occupied by each detonation. Figure 8 shows the joint probability density functions (jPDF) of the pressure and temperature of the reactants inside the front, i.e., in grid cells, in which the mass fraction of hydrogen $Y_{H_2} \geq 90\%$ of the upstream value. We focus on the state of the reactants rather than the entire detonation volume, since it is this shocked fresh mixture which controls ignition in the front, and thus its entire structure and dynamics. All joint PDF were collected over one flow-through time $\tau_{ftt} \equiv L/D_{CJ}$, where $L$ is the channel width, and they combined 200 individual flow snapshots. The cases shown with the yellow-red color map are in agreement with the experiments, while the cases shown with the blue color map represent detonations at atmospheric conditions, in which the cell size differs from the experimental values.

Overall, after passing the leading shock, reactants are present over a wide range of pressures and temperatures [18, 40]. Such conditions represent both singly shocked states along the primary shock adiabat (solid black and red lines), as well as multiply, in particular doubly, shocked states, which lie above the primary adiabat between the limiting secondary adiabats (dashed and dotted lines). The speed of the primary shock corresponding to each of these secondary adiabats is indicated in the figure. Note that the secondary adiabat, which corresponds to the underdriven primary shock speeds $< D_{CJ}$ (dashed lines in Fig. 8), marks the upper boundary of the region of doubly-shocked states. There is also a large region of states below the primary shock adiabat, which are characterized by lower pressures. Note that large fan-like sections of the jPDF, which extend to low pressures and temperatures, are the states inside the leading numerical shock structure.[3]

The VN state, i.e., the state behind the shock in an idealized CJ detonation (white and green dots in Fig. 8), is often used to characterize the detonation structure. For example, induction length, $\Delta_i$, in the ZND solution corresponding to the VN state (dash-dot lines below the primary adiabats), is traditionally

[3] We did not exclude the computational grid cells interior to the numerical shock, which is typically spread over 3-4 cells. It is not possible to precisely define the criterion to exclude such cells, since the leading shock strength and orientation can vary significantly in the detonation front. Therefore, any threshold value of pressure, temperature, etc. chosen to provide such selection could fail to exclude some of the grid cells inside the shock or would unnecessarily exclude cells behind it.

used to predict key dynamical characteristics of a detonation, including the detonation cell size [115], minimum channel size [116], or critical tube diameter, using various phenomenological correlations [117]. We emphasize, however, that the VN state, as well as the associated metrics based on the ZND solution, may not necessarily represent the properties of a detonation. For example, consider the stoichiometric $H_2$/air mixture at 1 bar. At an upstream temperature of 300 K, $\Delta_i$ = 216 µm according to the FFCM-1 mechanism [70] (Table 3). When the upstream temperature increases to 650 K, $\Delta_i$ increases by 17% to 253 µm, although the experimental cell size decreases by a factor of $\approx 2.7$ from $\approx 1.21$ cm to $\approx 4.5$ mm [42, 45] (Table 2). In contrast, the computed cell size increases somewhat from $\approx 3$ to 4 mm. In other words, even in this canonical mixture, ZND induction length does not correlate with the direction in which $\lambda_{exp}$ changes, and thus, if used in the traditional phenomenological correlations [47], it would not be able to predict $\lambda_{exp}$.

Therefore, the overall structure of a detonation depends very sensitively on all thermodynamic states occupied by it, rather than just the VN state, or even the entire ZND solution. More specifically, ignition delays, $\tau_{ign}$, both at the under- and overdriven states on the primary adiabat and at the multiply shocked states, are equally responsible for creating the correct dynamics of the transverse waves, which in turn determine the structure and stability of the overall front.

With these considerations in mind, we next examine the effect of the upstream conditions on the thermodynamic state space occupied by a detonation. In a stoichiometric $H_2$/air mixture under atmospheric conditions (blue color map in Fig. 8a; cf. Fig. 7a), the pressure of the reactants inside the front is $\gtrsim$ 10 bar and the temperature range is $\approx 900-2100$ K (excluding states inside the leading numerical shock). When the upstream pressure is reduced to 0.25 bar (yellow-red color map in Fig. 8a; see Fig. 7b), the temperature range becomes somewhat narrower $\approx 1100-1900$ K, while the pressures inside the front decrease below $\approx 10$ bar. When the mixture is preheated to 650 K at 1 bar (Fig. 8b; cf. Fig. 2a), resulting pressures in a detonation drop compared to the 298 K case to $\approx 6-30$ bar, while the temperatures of the reactants become significantly higher in the range $\approx 1300-2200$ K. Such lower pressures, compared to the 298 K case, are due to the much lower detonation shock strengths at 650 K as a result of a higher upstream sound speed.

The main effect of changing mixture composition on the thermodynamic state space of a detonation is shown in Fig. 8c for the same lean mixtures ($\phi = 0.5$) as in Fig. 7c,d. While the primary and secondary adiabats remain almost the same, reducing $\phi$ from 1.0 to 0.5 causes the overall distribution to shift along the corresponding primary adiabat towards lower pressures and temperatures. In particular, under atmospheric conditions, the lean mixture results in lower $D_{CJ}$ and weaker shocks in the front. As a result, both pressures and temperatures inside the front decrease compared to the stoichiometric case, respectively, to $\approx 5-30$ bar and $\approx 700-1500$ K. Furthermore, increasing the upstream temperature of this lean mixture to 650 K keeps virtually the same pressure range inside the detonation while shifting the distribution to higher temperatures $\approx 1200-2000$ K.

This leads to two key conclusions. First, agreement or disagreement of the numerical detonation structure with the experiments cannot be ascribed simply to the effect of pressure, which one may be tempted to do based on the fact that such agreement is well established at low pressures (also cf. Fig. 8a). The lean atmospheric case (blue color map in Fig. 8c) occupies a range of pressures similar to the preheated cases at $\phi = 1.0$ (Fig. 8b) and 0.5 (Fig. 8c), yet they all produce very different cellular structures: while the atmospheric case differs drastically from the experimental observations, both preheated cases agree reasonably well with the experiments. Second, shifting the thermodynamic distribution to higher temperatures (albeit at moderate pressures) improves agreement with the experiments. It is not clear whether this would be the case at elevated upstream pressures, i.e., if the state space were to shift to pressures above $\sim 100$ bar for the same range of higher temperatures. To our knowledge, there are no published detonation cell data in the $H_2$/air mixtures at simultaneously elevated upstream pressures and temperatures.

Finally, a question could be raised regarding the sensitivity of the observed detonation structure to the chemical kinetics model. Figure S3 in the Supplementary Material shows comparison of the soot foils from the simulations of the highly unstable mixture 4 (lean, atmospheric $H_2$/air) performed using the FFCM-1 [70, 125] (Fig. S3a) and the UCSD chemical mechanisms [126] (Fig. S3b). It shows that the resulting structure is very similar in both calculations. Note also that simulations performed by Taylor et al. [18] in the stoichiometric, atmospheric $H_2$/air mixture 1 used a different mechanism by Burke et al. [113], yet those simulations produced cell sizes very similar to those reported here (Tables 1 and 2). Finally, Smirnov et al. [20] considered stoichiometric $H_2$/air detonations at atmospheric pressure and a range of upstream temperatures from 300 K to 650 K using several chemical mechanisms different from the FFCM-1, UCSD, and Burke et al. [113] mechanisms. Yet, that study reported that at 300 K detonation cells were even smaller than those obtained by us and Taylor et al. [18], while at 650 K, they observed close agreement with the experimentally measured cells, again in agreement with our results. This shows that for $H_2$-air mixtures, presented observations regarding the detonation structure are independent of the specific chemical kinetics model.

## 6. Thermochemical Model of Shocked Flows in $H_2$/air Mixtures

These results show that the thermodynamic *P-T* state space of a detonation alone does not explain why certain cases agree with the experiments and some do

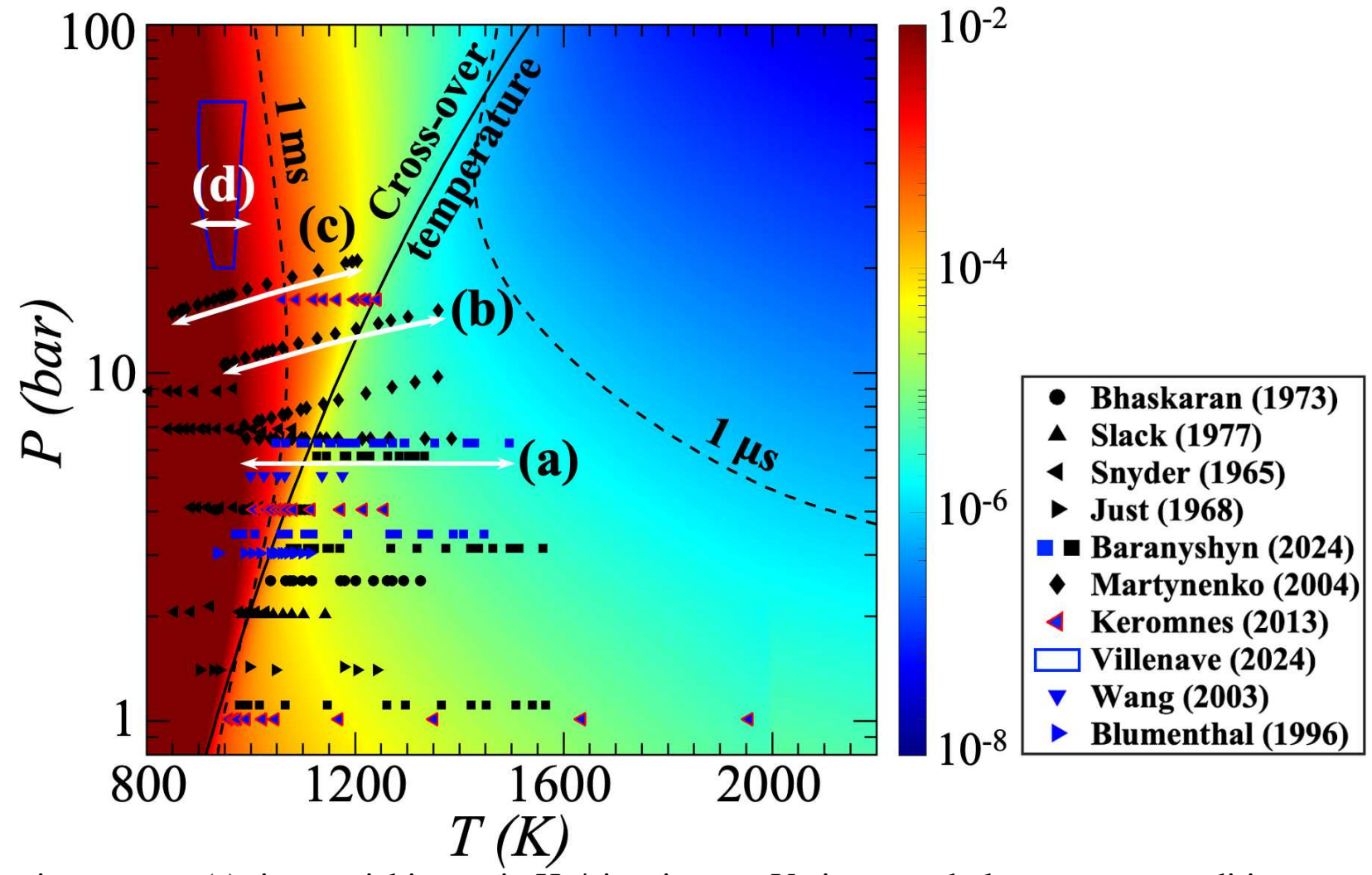


Fig. 9: Ignition times, $\tau_{ign}$ (s), in a stoichiometric $H_2$/air mixture. Various symbols represent conditions, at which $\tau_{ign}$ was experimentally measured in $H_2/O_2/N_2$ mixtures [104–106, 118–124]: black - stoichiometric; blue - lean; blue/red outline - lean with $N_2$ dilution different from air. White solid lines highlight the experimental conditions shown in panels (a)-(d) of Fig. 10.

not. The reason is that, ultimately, detonation dynamics critically depend not just on $P$ and $T$ inside the front, but rather on the response of chemical reactions to these conditions. This response can be quantified in terms of (i) the reactivity of the mixture, or more precisely the ignition delay time, $\tau_{ign}$, and (ii) its temperature sensitivity, namely how $\tau_{ign}$ varies with $P$ and $T$. In the context of detonations, such sensitivity is often characterized by the parameter [16, 88, 127]

$$\epsilon = \frac{1}{T_{VN}} \frac{\partial \ln \tau_{ign}}{\partial (1/T)}, \tag{1}$$

where $T_{VN}$ is the temperature at the VN state in a ZND solution of a CJ detonation, and the logarithmic derivative is calculated by perturbing the shock strength around $D_{CJ}$.

Figure 9 shows the map of the isobaric $\tau_{ign}$ in the $P$-$T$ space for the stoichiometric $H_2$/air mixture based on the FFCM-1 chemical model [70]. Note that this map does not vary significantly with the equivalence ratio, provided that the type and amount of the diluent remain the same. Therefore, this map could be also viewed to represent lean and rich mixtures.

Shocks with Mach numbers $Ma \sim 3$ can raise the mixture temperature to $\sim 800 - 1000$ K. At such temperatures, the ignition delay is $\gtrsim 1$ ms, and it varies little with pressure (see the 1 ms isoline in Fig. 9). As the temperature increases above $\sim 1000$ K (shock strengths $Ma \gtrsim 3.5$), the ignition time drops by 2-3 orders of magnitude to $\sim 10$ µs over just a few hundred K. Such high temperature sensitivity (and thus $\epsilon$) is often associated with the second extended explosion limit [128], which plays a central role in various explosive processes. The cause of this phenomenon is that at different temperatures and pressures, reactions proceed through different pathways, the relative importance of which varies with the thermodynamic conditions. After initial radical formation, the role of peroxide reactions gradually decreases at post-shock temperatures $\gtrsim 1000$ K, and ignition begins to be controlled by the exchange reactions of the form $A + BB + M$. Of these, the most dominant one is the H + $O_2$ + M $\rightleftharpoons HO_2^* \rightleftharpoons$ OH + O channel [129]. At lower temperatures, the 3-body sub-channel of this reaction dominates, in which the meta-stable vibrationally excited $HO_2$ is formed. At higher temperatures, however, it becomes difficult to stabilize this radical, which instead promotes the production of the chain branching OH and H radicals. Therefore, a rapid increase in mixture reactivity at higher temperatures is primarily controlled by the transition from the 3-body reaction, which inhibits chain branching, to the 2-body reaction, which drives thermal runaway. As a result, the zone of such high temperature sensitivity is approximately traced by the cross-over temperature (solid black line in Fig. 9a), above which the 2-body reaction dominates the 3-body reaction.

Figure 9 illustrates two important properties of the $H_2$ chemical kinetics, which are representative of virtually all modern reaction mechanisms. First, the cross-over temperature, and thus the region of high $\epsilon$, grows with pressure from $\approx 900$ K at 1 bar to $\approx 1500$ K at 100 bar due to the effect of pressure on the 3-body H + $O_2$ + M reaction.

Second, the distribution of ignition times above the cross-over temperature also differs qualitatively between low and high pressures. At pressures close to atmospheric, $\tau_{ign}$ drops to $\sim 10$ µs (green region in Fig. 9) within $\approx 400 - 500$ K once the balance shifts from the 3-body reaction to the 2-body reaction around $900 - 1000$ K. As the temperature increases further to 2000 K and above, $\tau_{ign}$ continues to decrease very slowly, remaining well above 1 µs.

At higher pressures $\gtrsim 10 - 15$ bar, the initial rapid drop in $\tau_{ign}$ from $\gtrsim 1$ ms to $\sim 10$ µs occurs somewhat faster than at lower pressures, namely over $\approx 300$ K. This drop, however, is located entirely to the left of the cross-over temperature, and, thus, it is controlled by the pressure response of the 3-body reaction itself. Furthermore, in contrast to low pressures, $\tau_{ign}$ continues to decrease rapidly with increasing temperature reaching 1 µs at $T \sim 1450$ K and only a few tens of nanoseconds at temperatures close to 2000 K. For example, in a detonation at $\phi = 1$ and atmospheric conditions, $\tau_{ign}$ varies by more than five orders of magnitude from $\sim 10^{-2} - 10^{-3}$ s at $\sim 1000$ K in the underdriven sections of the front to $\lesssim 10^{-7}$ s at $\sim 2000$ K in the doubly-shocked overdriven ones (cf. Fig. 8a). This dramatic increase in temperature sensitivity of the reactions at high pressures is represented by the 1 µs isoline, which shifts to lower temperatures with pressure, approaching the 1 ms isoline within $\approx 350 - 400$ K at $P \approx 30$ bar.

An immediate question can be asked: how realistic are such sub-µs ignition delays? Comparison of Figs. 8 and 9 shows that large parts of the thermodynamic state space of the detonations in the preheated mixtures (650 K) both at $\phi = 1$ and 0.5 fall within the region of the ultra-fast reactions to the right of the 1 µs isoline. Furthermore, in these two cases, $\tau_{ign} \lesssim 1$ µs even at the VN state. Despite this, both cases show good agreement with the experimental cell sizes. This suggests that the existence of such sub-µs ignition delays may not necessarily be unphysical. However, it is less clear (i) how small $\tau_{ign}$ can be, i.e., if its values can indeed approach $\sim 10^{-8}$ s or less at higher $P$ and $T$, and (ii) what region of the $P$-$T$ space is characterized by the sub-µs ignition delays, i.e., what the shape of the 1 µs isoline is. Note that at higher pressures, a large part of the thermodynamic state space of the detonation at $\phi = 1$ without mixture preheating (298 K and 1 bar), including its corresponding VN state, also ends up in the region with $\tau_{ign} < 1$ µs. However, in this case, pronounced disagreement with the experimental cell sizes does not allow one to conclude whether such short ignition delays at higher pressures are realistic or not.

There is another important distinction between the two cases in Fig. 8, which disagree with the experiments ($\phi = 0.5$ and 1.0 under atmospheric conditions), and the other three cases, in which correct cells are obtained. In particular, low-pressure $H_2$/air detonations, as well as detonations in the preheated mixtures (Fig. 8), almost entirely occupy states to the right of the cross-over temperature. Such states, including the corresponding VN state (see values of $\epsilon$ in Table 3), are characterized by a much lower temperature sensitivity of $\tau_{ign}$. As was shown in previous studies [16, 88], lower values of $\epsilon$ produce more regular detonations, which is indeed reflected in the soot foils in Figs. 2a,b and 7b,d.

In contrast, the lower-temperature part of the state space of the detonation in a stoichiometric mixture under atmospheric conditions (Fig.8a) traverses the high-$\epsilon$ region close to the cross-over temperature, although the VN state lies to the right of it (in the sub-µs region). When $\phi$ is reduced to 0.5 (Fig. 8c), the entire state space shifts to lower temperatures with the predominant majority of states falling to the left of the cross-over temperature in the region of very high temperature sensitivity, with $\epsilon = 36$ at the VN state. Such high $\epsilon$ is the direct cause of a highly unstable detonation structure with extremely small cells (Fig. 7c).

These observations show that detonations under varying upstream conditions and in different mixtures probe different regions of the $P$-$T$ space, in which the reaction kinetics exhibits qualitatively different behavior. Both the magnitude of $\tau_{ign}$, and its temperature sensitivity, directly translate into the structure and dynamics of a detonation. Thus, agreement, or lack thereof, of the computed structure with experimental observations serves as an indirect probe of the accuracy of the description of the reaction kinetics.

It is important to emphasize that relative importance of $\tau_{ign}$ and $\epsilon$, i.e., the derivative of $\tau_{ign}$ with temperature (and pressure), can be different in high- and low-$\epsilon$ mixtures. For instance, consider stoichiometric and lean $H_2$/air mixtures 1 and 4. The joint PDF show that the state space of the lean mixture shifts to the region of much larger ignition delays, yet the numerically calculated cells become much smaller. This suggests that it is $\epsilon$, which primarily controls the structure, rather than $\tau_{ign}$, since if it were only the latter, then cell size would increase relative to the stoichiometric composition. Instead, $\epsilon$ increases considerably causing much smaller and much more irregular cells. Ultimately, however, $\epsilon$ is determined by the distribution of $\tau_{ign}$ in the $P-T$ space. Therefore, to establish whether a particular mixture is characterized by high or low $\epsilon$, one must constrain the chemical properties for the relevant conditions.

This raises the question whether direct, and more quantitative, evidence exists of the accuracy of modern thermochemical models of hydrogen oxidation across a wide range of pressures and temperatures formed in strongly shocked flows? Symbols in Fig. 9 show existing experimental measurements of ignition delays in the $H_2/O_2/N_2$ mixtures from the last half century. Here we collected published data only in air- (or air-like) compositions. There exists a large body of measurements of $\tau_{ign}$ at elevated pressures in mixtures with very high Ar dilution (see detailed review by Schultz and Shepherd [130]). Such data, however, cannot be reliably used to validate the reaction kinetics in more realistic compositions in pure $O_2$ or air due to the strong dependence of both $\tau_{ign}$ and its temperature sensitivity on the amount and type of a diluent (compare Fig. 9 with the $P$-$T$ map of $\tau_{ign}$ for a heavily Ar-diluted $H_2/O_2$ mixture in Fig. S4 in the Supplementary Material). In particular, large amounts of monoatomic diluents significantly change both the thermodynamic properties of the mixtures (namely $\gamma$), as well as the reaction kinetics, in particular 3-body reactions, which depend sensitively on the third-body efficiency factors. These factors, along with a much lower overall exothermicity of a highly diluted mix-ture, result in ignition delays, which can be orders of magnitude larger than in more realistic $H_2$/air compositions, thus making direct comparisons with the chemical dynamics in air problematic.

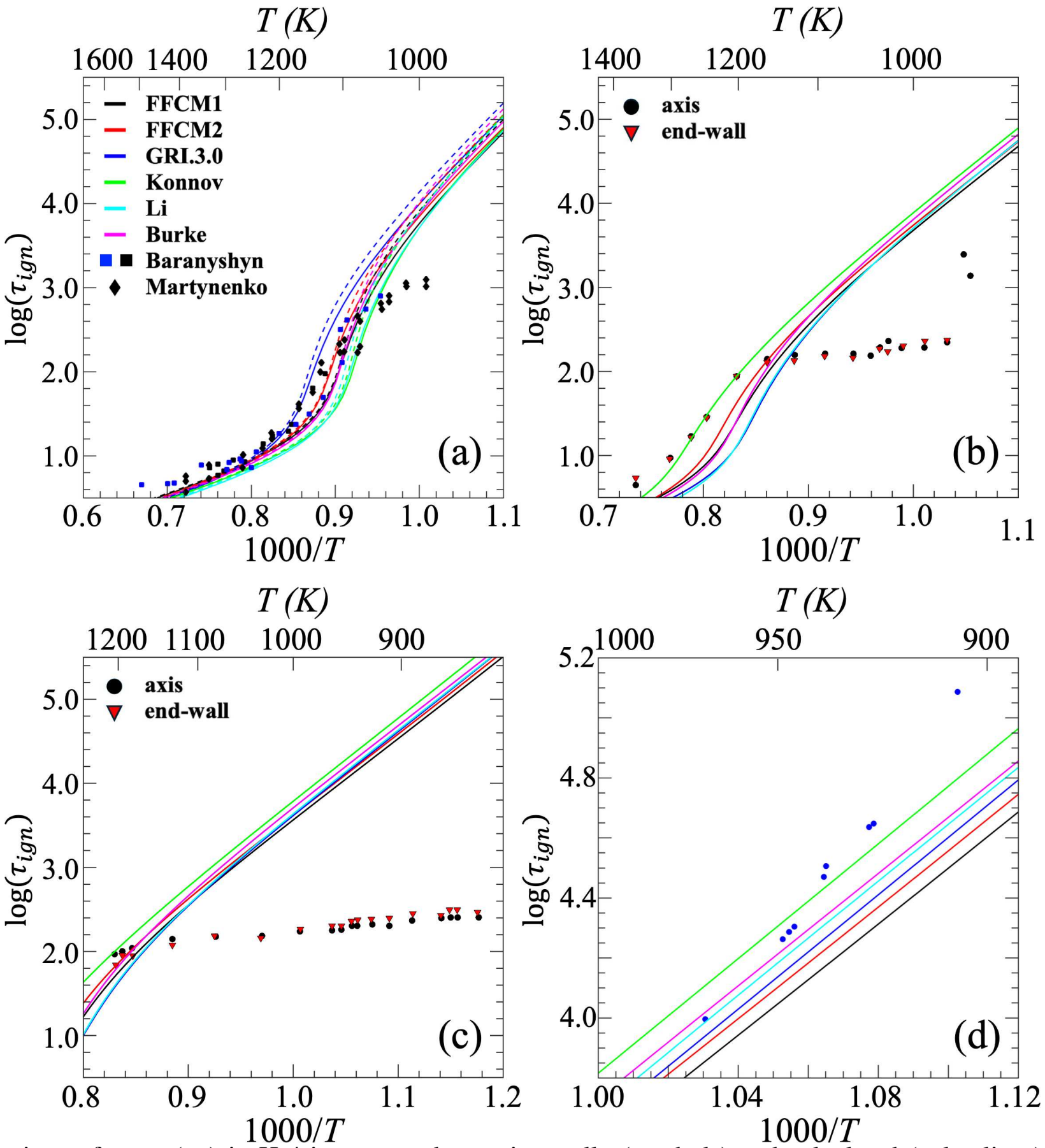

Fig. 10: Comparison of $\tau_{ign}$ (µs) in $H_2$/air measured experimentally (symbols) and calculated (color lines) using different chemical mechanisms listed in panel (a) [70, 72, 110–113]. (a) Shock-tube data from Refs. [104, 105]. Black symbols: stoichiometric mixtures; blue symbols: lean mixtures. The constant-volume ignition delays for different chemical mechanisms at $\phi = 1.0$ (solid) and $\phi = 0.5$ (dashed) are computed at pressures of 5.7 and 6.2 bar, respectively. (b,c) Shock-tube data from Ref. [105] obtained on the axis and at the end-wall for initial densities of 2.8 (b) and 4.39 kg/m$^3$ (c), respectively. (d) Rapid compression machine data from Ref. [106] for $\phi = 0.5$ at 30 bar. Thermodynamic conditions for the experimental measurements in each panel are shown in Fig. 9, in which the data set corresponding to each panel (a), (b), (c), and (d) is marked accordingly and is highlighted with a solid white line.

The experimental conditions marked in Fig. 9 show two key limitations of the existing ignition delay data. First, the vast majority of measurements were performed at lower pressures $\lesssim 10$ bar, with only a handful of datasets obtained at higher pressures in the last 20 years [105, 106, 122]. To our knowledge, no shock-tube measurements exist in $H_2$/air for pressures $\gtrsim 20-30$ bar. Second, at higher pressures, data exist only for temperatures $\lesssim 1200-1500$ K. As a result, existing measurements are limited by ignition delays $\gtrsim 5-10$ µs.

Figure 10 compares the experimentally measured ignition delays with the predictions of several leading chemical mechanisms [70, 72, 110–113]. Here, we focus on four data sets obtained at higher pressures, which are the most relevant to the conditions present in $H_2$ detonations discussed above (Fig. 8).

Figure 10a shows $\tau_{ign}$ in the stoichiometric and lean $H_2$/air mixtures obtained by Baranyshyn et al. [104] and Martynenko et al. [105] at pressures close to $\approx 6$ bar and temperatures in the range $\approx 1000-1500$ K (cf. points marked with the lowest white line labeled (a) in Fig. 9). These conditions well represent the thermodynamic states in a detonation at $\phi = 1$ and low pressure of 0.25 bar (293 K; Fig. 8a).

At higher temperatures $\gtrsim 1200$ K, theoretical and experimental ignition delays are in close agreement, despite some scatter in the experimental and theoretical values. In particular, the values of $\tau_{ign} \sim 5-10$ µs are predicted and indeed observed. Chemical models also capture reasonably well the location of the sharp increase in temperature sensitivity around 1100 K, even though the difference between individual models can be $\approx 50-70$ K. Finally, at the lowest temperatures approaching 1000 K, chemical models appear to deviate from the experimental measurements, predicting much larger $\tau_{ign}$ and higher temperature sensitivity. Despite this, overall agreement demonstrates that, at these relatively low pressures, modern thermochemical models are indeed able to capture the key features of the reaction kinetics. This agreement translates into accurate predictions of the detonation structure at low pressures, well documented in numerous studies [16, 88, 90].

Note that these existing $\tau_{ign}$ data barely reach the lowest temperatures present in a detonation in the preheated stoichiometric mixture (Fig. 8b). At the same time, as noted above, good agreement of the detonation structure obtained in simulations with experiments provides indirect evidence of the validity of the existing thermochemical models also at temperatures $\gtrsim$ 1500 K and pressures $\lesssim 20 - 30$ bar. In this regard, detonations can represent an alternative to direct shock-tube measurements, which could allow one to probe the high-pressure/high-temperature reaction kinetics, although any such indirect assessments must be treated with significant caution.

The situation is more complex at higher pressures, which are representative of detonations at atmospheric conditions. Figure 10b,c shows experimental data from Martynenko et al. [105] in a stoichiometric $H_2$/air at pressures $\approx 10 - 20$ bar and temperatures $\approx 850 - 1350$ K. These conditions probe the lower $P$-$T$ part of the thermodynamic state space in a detonation in a stoichiometric mixture (Fig. 8a) and most of the state space in a detonation in a lean mixture (Fig. 8c), both at atmospheric conditions. In particular, they span the critical region of high temperature sensitivity, which controls the size and regularity of the detonation cells. Therefore, in principle, these data would be able to reveal any inaccuracies that may exist in the theoretical ignition delay values. Figures 10b,c show that the experimental measurements agree with the theoretical predictions only at the highest temperatures $\gtrsim$ 1150 K, at which $\tau_{ign} \lesssim 100$ µs. At lower temperatures, the measured $\tau_{ign}$ appears virtually insensitive to temperature, causing theoretical values to exceed the experimental points by several orders of magnitude.

Such stark disagreement, however, should not be viewed as evidence of the inaccuracy of the thermochemical model, but rather as the result of the limitations in shock-tube measurements. Short ignition delays likely do not represent homogeneous autoignition at the nominal reflected-shock temperature and pressure, but rather are the result of boundary-layer effects, localized ignition, and flame propagation [105]. Other effects can also limit measurements of long ignition delays $\gtrsim 0.1 - 1$ ms in shock tubes, especially at high pressures (see Refs. [131, 132] for a more detailed discussion).

Finally, Fig. 10d shows sample rapid compression machine (RCM) measurements of $\tau_{ign}$ at even higher pressures $\sim 20 - 50$ bar, but for a relatively narrow range of low temperatures $\approx 900 - 1000$ K, which corresponds to much longer ignition delays $> 10$ ms. Reasonable agreement with theoretical models can be seen to further suggest that the disagreement present in the data in Figs. 10b,c is due to experimental uncertainties rather than chemical model inaccuracies. At the same time, at the lowest temperatures approaching 900 K, all chemical mechanisms appear to underestimate $\tau_{ign}$ by a factor of $\approx 3$, depending on the model.

## 7. Potential Role of Vibrational Non-equilibrium in Strongly Shocked Flows

These results show that at lower pressures $\lesssim 10$ bar, at which modern thermochemical models have been developed and experimentally validated using direct ignition delay measurements, the computed detonation structure agrees closely with experimental results. At the same time, at elevated pressures $\gtrsim 10 - 20$ bar, there are only very limited direct ignition delay measurements in $H_2$/air mixtures, which are reliable enough to allow one to probe the accuracy of the existing thermochemical models for temperatures $\sim 800 - 2000$ K, which are dynamically important in detonations under atmospheric (or higher-pressure) conditions. Furthermore, uncertainties of the 2-body $H+O_2$ reaction were directly investigated only at pressures close to atmospheric [133] in heavily Ar-diluted mixtures, while uncertainties of the 3-body reaction were analyzed at $\approx 12$ bar and temperatures $1200 - 1400$ K for the $N_2$ collider [134]. No systematic investigation of these key reactions has been conducted in strongly shocked, high-pressure air mixtures. As a result, one cannot rule out the limitations of the existing thermochemical models as the source of the disagreement discussed above between the computed detonation structure and experiments.

One such potential limitation is the fact that virtually all existing $H_2$ chemistry models are built on the assumption of thermal equilibrium. This assumption, however, is violated immediately behind strong shocks, since gas still has low rovibrational energy of the upstream flow and very high translational temperature. While rotational degrees of freedom typically relax more quickly [135], vibrational relaxation proceeds more slowly. Therefore, a question naturally arises whether vibrational non-equilibrium (VNE) can *qualitatively* alter the dynamics of strongly shocked reacting flows.

The importance of VNE in strong air shocks for aerothermochemistry has long been recognized and actively studied in the hypersonic community [136, 137]. Numerous experiments have shown that VNE significantly affects post-shock temperatures and composition through a two-way coupling between the finite rate of vibrational–translational energy exchange and the dissociation rate [138–140].

In chemically reacting flows, however, reactions typically occur at much higher pressures and much lower temperatures than in hypersonic flows. Thus, the role of VNE may not necessarily be the same. In detonative flows, potential importance of the flow non-equilibrium has been suggested since the 1950's, when shock-tube experiments showed that the induction period increases with the vibrational relaxation time of $O_2$ [141]. The proximity of the chemical and inert vibrational relaxation timescales was first suggested by Taylor et al. [18] as a possible cause of the discrepancy between the calculated and experimental cell sizes in a stoichiometric $H_2$/air detonation [26]. Since then, several studies have developed a generalized ZND solution for the 1D detonations [142, 143]. Uy et al. [142] showed that VNE can indeed increase the induction length, which also agrees with the analysis of Fievet et al. [144] in the context of VNE that arises from the rapid fuel/oxidizer mixing in scramjets. The fact that $\tau_{ign}$ becomes larger in the presence of VNE suggests that this should also lead to an increase in cell size. In fact, such an increase was observed in 2D simulations [27, 28], although the effect was modest with an increase in cell size between 30 – 60% [27] and a factor of two [28].

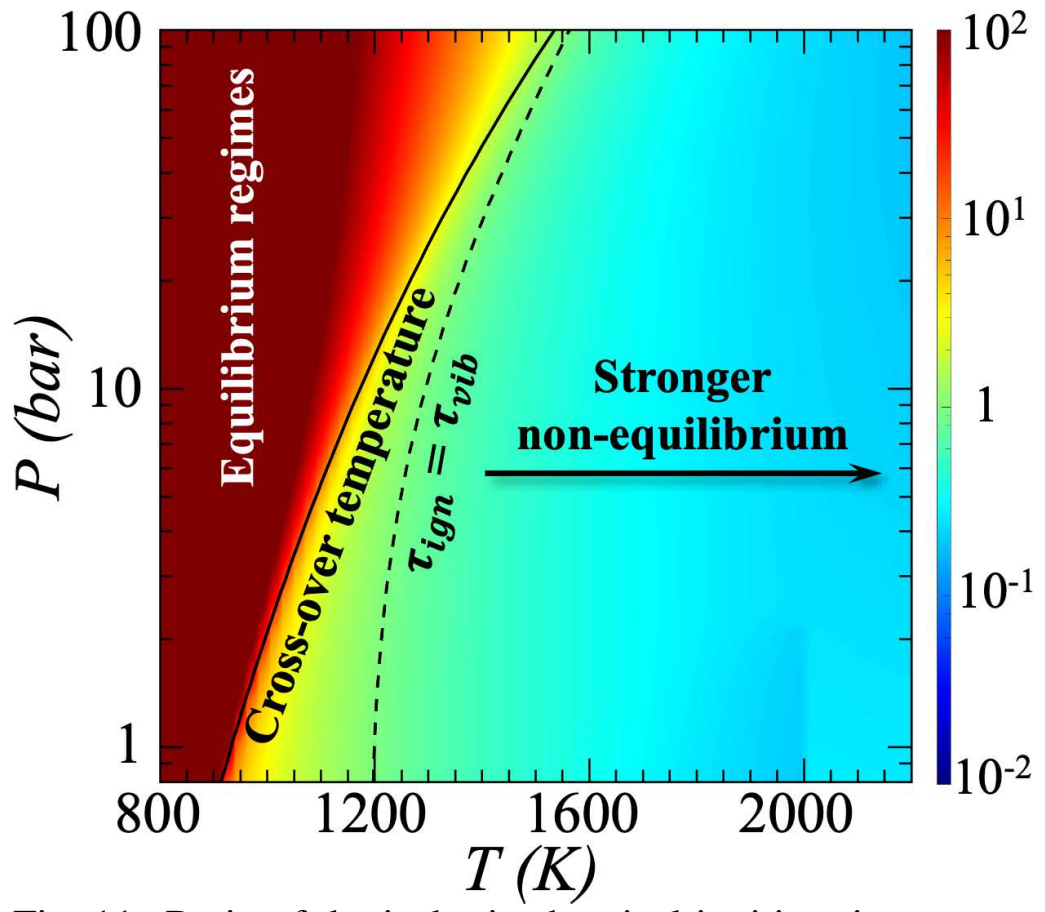

Fig. 11: Ratio of the isobaric chemical ignition time, $\tau_{ign}$, to the vibrational relaxation time, $\tau_{vib}$, of $N_2$ in the stoichiometric $H_2$-air mixture.

Figure 11 further illustrates the rationale behind the suggested importance of VNE in shocked reacting flows. It shows the ratio of the isobaric chemical ignition time, $\tau_{ign}$, to the inert vibrational relaxation time, $\tau_{vib}$, of $N_2$ in the stoichiometric $H_2$/air mixture (also see Taylor et al. [26]). The $\tau_{vib}$ is calculated using the Landau-Teller model with the characteristic relaxation time given by the classical Millikan-White approximation [26, 145]. Note, we consider $N_2$ since it has the slowest relaxation time in a $H_2$/$O_2$/$N_2$ mixture, compared to $O_2$ and $H_2$. The temperature in Fig. 11 is the equilibrium temperature, at which $\tau_{ign}$ is calculated using the FFCM-1 mechanism [70]. The dashed line marks the conditions, at which $\tau_{vib} = \tau_{ign}$. At pressures and temperatures to the right of this line, Fig. 11 suggests that the physicochemical model is physically inconsistent since it assumes equilibrium conditions, while vibrational relaxation would occur on similar or much longer timescales [26]. It is important to emphasize though that direct comparison of the inert vibrational relaxation time scale with the ignition time based on the equilibrium kinetics, such as shown in Fig. 11, may not account for the full complexity of the coupling between reactions and vibrational relaxation.

Comparison of this analysis in Fig. 11 with the thermodynamic state space of the $H_2$/air detonations (Fig. 8) leads to somewhat paradoxical conclusions. Thermodynamic states of detonations in preheated mixtures (both at $\phi = 1.0$ and 0.5) lie in the region, which should, in principle, be characterized by strong non-equilibrium (to the right of the $\tau_{vib} = \tau_{ign}$ line in Fig. 11). Yet both cases are in agreement with experiments despite the fact that they use equilibrium kinetics. Furthermore, as discussed above, at lower pressures, current equilibrium kinetics is in agreement both with the direct ignition delay measurements and macroscopic detonation simulations.

In contrast, detonations at atmospheric conditions either partially (at $\phi = 1.0$) or almost entirely (at $\phi = 0.5$) occupy states to the left of the $\tau_{vib} = \tau_{ign}$ line, at which thermal equilibrium should re-establish faster than the characteristic chemical timescale. However, in both cases, existing thermochemical models produce an incorrect detonation structure. In fact, as one moves from the stoichiometric to a lean composition, shocks in a detonation front become weaker. This would reduce, on one hand, the post-shock translational temperature and thus the degree of non-equilibrium, and on the other, increase the chemical timescales, thus pushing the detonation further into the equilibrium regime. Instead, the disagreement of simulations with experiments becomes only larger.

These results, along with the observations made in prior studies, which attempted to incorporate the non-equilibrium effects in the thermochemical model, paint a complex, and at present contradictory, picture of the potential role of VNE in shocked reacting flows. Consequently, the following three hypotheses can be proposed.

(i) VNE is important at all pressures. At lower $P$, at which existing chemistry is anchored by the experimental data, reaction rates implicitly account for the VNE effects. However, when extrapolated to higher pressures, reaction rates become progressively less accurate.

(ii) At lower pressures, the equilibrium assumption is indeed valid, and the existing thermochemical models are physically sound. However, at higher pressures, VNE becomes more important and must be properly accounted for.

(iii) Finally, the equilibrium assumption is valid at all pressures, and there exist some other causes of the discrepancy between the detonation simulations and experiments.

Establishing which of these hypotheses is correct, in our view, is central to the understanding both of the physics of detonations and, more broadly, of the physicochemistry of high-speed reacting flows.

We emphasize that even under the conditions of thermal equilibrium, the vibrational state of the multi-atomic colliders plays a defining role in both the dissociation and exchange reactions. For example, in the 2-body $A + BB$ exchange reactions, the transient complex ($ABB^*$) may collide with a third species, forming products different from those formed via the recombination channel ($ABB$). These non-thermal reactions require direct treatment of the vibrational energy, as shown by Burke and Klippenstein [129]. The initial vibrational energy of BB affects the termolecular reaction rates, highlighting the coupling between the finite-rate energy transfer and 3-body kinetics. Although that original study used a master equation model to capture the non-equilibrium effects, the reported rates apply only to flows in thermal equilibrium since low-speed near-isobaric flames, rather than shocked flows [146], were considered.

Most prior studies, which analyzed the effect of VNE in detonative flows, relied on the multi-temperature (MT) formalism, originating with Park's classical model of thermochemical non-equilibrium in high-temperature gases [136, 155]. Developed for high-altitude atmospheric re-entry and calibrated to match the post-shock composition [156], that approach was supported by shock-tube experiments using $N_2$–Ar or $O_2$–Ar mixtures shocked to 6000–9000 K at sub-atmospheric pressures. Detonation fronts, however, operate under much lower temperatures and much higher pressures (Fig. 8), at which exchange and recombination processes play a major role, making the validity of the MT models questionable. Although coupling of the MT models with several key reactions of the equilibrium chemistry has suggested the importance of VNE [27, 28], the MT approach is insufficient to resolve discrepancies between simulations and experimental observations of detonations.

The MT models are not the only framework explored in reacting-flow studies. Several works developed a master equation description, explicitly resolving the lowest vibrational states of $O_2$ and $H_2$ and assessing their impact on the detonation structure [157, 158]. These studies showed that the master equation model yields more accurate results than the MT models [158]. However, the state-resolved approach improves accuracy only when supplied with state-specific rate constants of comparable fidelity, which are typically obtained from trajectory simulations on the relevant potential energy surfaces. VNE effects in detonations have also been demonstrated using other methods, e.g., discrete Boltzmann method [107] and Direct Simulation Monte Carlo [108].

Finally, recently a number of experimental studies have attempted to quantify the VNE and assess its role in the detonation dynamics at sub-atmospheric pressures [159–162]. Obtained experimental evidence largely indicates that VNE does not have a significant effect. This observation is consistent with the fact that numerical simulations, which are performed at similar lower pressures using existing equilibrium chemical kinetics mechanisms, are also generally in agreement with the experimental measurements of the detonation structure.

## 8. Thermochemical Model of Shocked Flows in Hydrocarbon Fuels

The analysis in §§5-7 focused on the $H_2$ mixtures. At the same time, we discussed in §3 that detonation simulations in various hydrocarbon fuels, which use state-of-the-art chemical mechanisms, also show significant disagreement with the experimentally observed detonation structure. Therefore, it is important to discuss for representative hydrocarbons, albeit briefly, the same two central questions, namely (i) what the chemistry response is at different pressures and temperatures, and (ii) how well modern hydrocarbon chemistry is constrained by direct experimental measurements in the detonation-relevant regimes. Here we focus on the same fuels, for which soot foils were shown in Fig. 5, namely $CH_4$, $C_2H_2$, and JP-10 (mixtures 6, 9, and 10 in Table 2).

Figures 12a-c show $\tau_{ign}$ for these three fuels based on the FFCM-1 [73, 74] (panel a), FFCMy (panel b), and FFCM-2 [72] (panel c) chemical mechanisms. In each panel, color circles mark the VN state in a CJ detonation, while the upward and downward sloping solid lines passing through the VN state indicate the primary shock adiabat and the ZND solution, respectively. Dashed and dash-dotted lines starting on the primary adiabat represent secondary doubly-shocked adiabats corresponding to the primary shock speed of $0.7D_{CJ}$ and $D_{CJ}$, respectively. The region bounded by the primary and secondary shock adiabats approximately outlines the thermodynamic states in detonations in mixtures 6, 9, and 10 (cf. Fig. 8).

In $CH_4$ (mixture 6 in Table 2), the distribution of $\tau_{ign}$ is qualitatively simpler than in $H_2$ (Fig. 9). The ignition delays show pressure dependence, with their values decreasing with increasing pressure at a constant temperature. At the same time, the temperature sensitivity of $\tau_{ign}$ does not exhibit the same strong increase with pressure seen in hydrogen. In particular, the 1 ms and 1 µs isolines remain almost parallel to each other across the full pressure range, simply shifting to lower temperatures at higher pressures.

In a heavy hydrocarbon fuel, namely JP-10 (mixture 10 in Table 2), the distribution of $\tau_{ign}$ is qualitatively similar to that of methane (Fig. 12c). In particular, $\tau_{ign}$ also exhibits pressure dependence, becoming smaller as pressure increases at a given temperature. Furthermore, individual isolines are also almost parallel (see 1 ms and 1 µs isolines) throughout the full pressure range, and their slope is similar to that in $CH_4$. The main difference lies in the ignition delay values, which are substantially smaller than in $CH_4$ with the 1 ms and 1 µs isolines shifted to lower temperatures by $\approx 200 - 300$ K.

In both $CH_4$ and JP-10, the locations of the primary and secondary adiabats, as well as of the VN state, in the detonations at atmospheric conditions show that although ignition delays vary by almost three orders of magnitude between the over- and underdriven sections of the front, they remain above a µs. Therefore, unlike $H_2$, the question of the physical realism of the ultra-short sub-µs ignition delays becomes relevant to these fuels only at the upstream pressures and temperatures well above atmospheric.

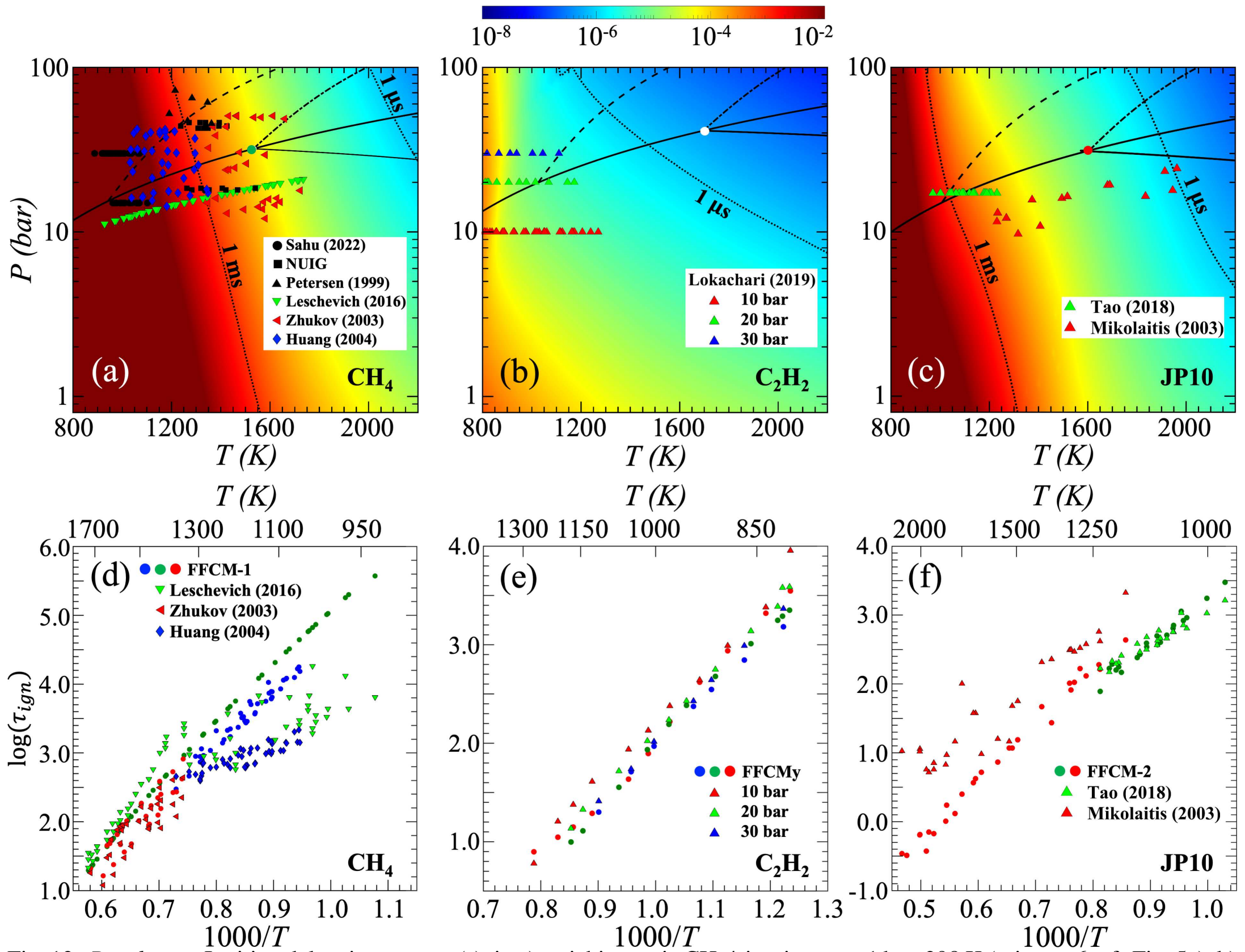

Fig. 12: *Panels a-c:* Ignition delay times, $\tau_{ign}$ (s), in a) stoichiometric $CH_4$/air mixture at 1 bar, 298 K (mixture 6; cf. Fig. 5a), b) rich ($\phi$ = 3) $C_2H_2$/air at 1 bar, 293 K (mixture 9; cf. Fig. 5b), and c) rich ($\phi$ = 1.4) JP-10/$O_2$/$N_2$ mixture at 1 bar, 353 K (mixture 10; cf. Fig. 5c). Various symbols represent conditions, at which $\tau_{ign}$ was experimentally measured in the stoichiometric a) $CH_4$/air [147–152], b) $C_2H_2$/air [153], and c) JP-10/air [71, 154]. Upward sloping solid lines: primary shock adiabat; dashed and dash-dotted lines: secondary adiabats corresponding to the primary shock speeds $0.7D_{CJ}$ and $D_{CJ}$, respectively; downward sloping solid lines: ZND solutions. Color circles mark the VN state. *Panels d-f:* Comparison of the experimentally measured and theoretically predicted $\tau_{ign}$ (µs) for the conditions indicated with the corresponding color symbols in panels (a)-(c). Circles show theoretically calculated $\tau_{ign}$ values (using FFCM-1, FFCMy, and FFCM-2 chemical mechanisms) for the experimental conditions marked with symbols of the same corresponding color.

Figure 12b shows the distribution of $\tau_{ign}$ in $C_2H_2$/air (mixture 9 in Table 2). In acetylene, the chemistry response to varying thermodynamic conditions is qualitatively different from methane and JP-10. At low temperatures $\lesssim 800-1000$ K and pressures greater than a few bar, $\tau_{ign}$ is virtually independent of pressure. At the same time, at higher temperatures, ignition delays decrease rapidly with increasing pressure, which can be seen in the rapid shift of the 1 µs isoline to lower temperatures. As a result, the VN state, as well as a large region of the thermodynamic states of a detonation in atmospheric $C_2H_2$/air, have ignition delays significantly below 1 µs, with the most extreme overdriven and doubly shocked states reaching $\tau_{ign} \lesssim 10^{-7}$ s. Overall, at detonation-relevant pressures $\gtrsim 10$ bar and temperatures $\gtrsim 1200$ K, $C_2H_2$-air mixtures have both very short ignition delays, which reflect high reactivity of this fuel, and relatively low temperature sensitivity of $\tau_{ign}$ with $\epsilon \approx 4$ at the VN state (Table 3).

Similar to hydrogen, it is instructive to compare these theoretical ignition delays with the actual direct experimental measurements in air or near-air compositions. Again, we are primarily interested in conditions formed behind strong shocks, i.e., high pressures $\gtrsim 10$ bar and temperatures $\gtrsim 1000$ K. In $CH_4$/air, there is a considerable amount of experimental data at high pressures $\approx 10-70$ bar and temperatures as high as $\approx 1750$ K. Various symbols in Fig. 12a show that, unlike $H_2$/air, existing data cover a large part of the thermodynamic state space of the atmospheric $CH_4$/air detonations. In Fig. 12d, we show measured values of $\tau_{ign}$ from three representative data sets, namely by Leschevich et al. [150], which spans temperatures from 900 to 1700 K at pressures somewhat below the primary adiabat, as well as the data sets by Zhukov et al. [151] and Huang et al. [152], which jointly cover a similar temperature range as the data by Leschevich et al. [150] but which were obtained at higher pressures (cf. the same color symbols in Fig. 12a). Finally, Fig. 12d also shows theoretical values obtained with the FFCM-1 mechanism [70] at the same $P$ and $T$ from each experimental measurement (circles of the same color as the corresponding experimental dataset).

Figure 12d shows that in $CH_4$/air at higher temperatures $\gtrsim 1300$ K, theoretical and experimental values agree, despite the considerable scatter in the experimental data, in particular between the data sets of Leschevich et al. [150] and Zhukov et al. [151]. This is the region to the right of the 1 ms isoline in Fig. 12a closer to the VN state. Note that in this region, the FFCM-1 mechanism captures not only the values of $\tau_{ign}$, but more importantly its slope, or the temperature sensitivity $\epsilon$.

At lower temperatures $\lesssim 1300$ K, which correspond to the underdriven sections of the front, the agreement with the experimental ignition delay measurements in $CH_4$/air is considerably worse. In particular, the data set of Zhukov et al. [151] indicates that $\tau_{ign}$ varies little as the temperature decreases to $\approx 1000$ K remaining close to $\sim 1$ ms. In contrast, theoretical values continue to grow with decreasing temperature almost with the same slope as at higher temperatures. As a result, close to 1000 K, theoretical ignition delays are larger than the experimental values by more than a factor of ten. Furthermore, theoretical temperature sensitivity of $\tau_{ign}$ in this range is also much higher than the measured one. The data of Leschevich et al. [150] in this temperature range have a much larger scatter compared to those of Zhukov et al. [151]. However, it generally agrees with the trend seen in the latter dataset with values of $\tau_{ign}$ significantly shorter than the theoretically predicted ones and much lower effective $\epsilon$.

Unlike methane, there are far fewer experimental high-pressure ignition delay data in $C_2H_2$/air. The color symbols in Fig. 12b show thermodynamic conditions, at which $\tau_{ign}$ was measured at 10, 20 and 30 bar by Lokachari et al. [153]. These conditions barely reach the most underdriven states in the atmospheric $C_2H_2$/air detonation, and thus they do not constrain the accuracy of the existing thermochemical models. Nevertheless, Fig. 12e shows experimentally measured and theoretically predicted $\tau_{ign}$ at these pressures and temperatures. In the range of conditions surveyed, ignition delays are virtually independent of pressure at a given temperature, and, overall, theoretical and experimental values agree.

Finally, Fig. 12c shows thermodynamic conditions, at which $\tau_{ign}$ was measured in the stoichiometric JP-10/air mixture by Tao et al. [71] and Mikolaitis et al.[4][154]. Both sets of measurements complement each other, effectively spanning the full range of temperatures present in an atmospheric detonation $\approx 950-2000$ K, albeit at somewhat lower pressures below the primary shock adiabat.

Figure 12f compares experimental measurements with the theoretically predicted values of $\tau_{ign}$, which were obtained at the same pressures and temperatures. Unlike $CH_4$, for JP-10, the best agreement between experimental and theoretical values exists at the lower temperatures $\approx 950-1250$ K surveyed by Tao et al. [71], which correspond to the most underdriven states in a detonation. Here, both $\tau_{ign}$ and its temperature sensitivity are correctly captured by the FFCM-2 mechanism. At higher temperatures, however, theoretical and experimental values diverge, even accounting for the significant scatter in the measured values. Predicted values appear to be systematically smaller than measured ones, with theoretical $\tau_{ign}$ reaching $\approx 0.3$ µs at $T \approx 2000$ K, while the experimental values remain closer to 10 µs. This results in a larger $\epsilon$ across virtually the entire temperature range of an atmospheric detonation in JP-10 with $\epsilon \approx 11$ at the VN state, somewhat lower than for methane, but much higher than in acetylene (Table 3). Such higher temperature sensitivity of $\tau_{ign}$ in both methane and JP-10 predicted by modern chemical models relative to the experimental measurements may, at least partially, explain extremely small and highly irregular detonation cell structure seen in Figs. 5a,c.

## 9. Discussion and Conclusions

Any theory is ultimately judged by its ability to explain existing observations and make verifiable predictions about future observations. Based on this epistemological criterion, what can be said about the state of our understanding of high-speed combustion? Despite the monumental advances made over the last century in thermochemistry, fluid dynamics, and computational methods, modern combustion theory of high-speed reacting flows is unable to reliably predict the structure and dynamical properties of the most canonical example of such flows - detonations. It could be argued that producing an accurate, robust, and predictive computational model of a freely propagating detonation constitutes one of the grand challenges of modern computational fluid dynamics. On the basis of the available experimental and numerical data, which we attempted to summarize in this paper, the following key observations can be made.

First and foremost, modern computational models based on the state-of-the-art chemical kinetics are not always wrong. In fact, in $H_2$/air mixtures, they can reproduce the structure and dynamics of detonations with good accuracy in a wide range of mixtures, from stoichiometric to lean, and in a wide range of conditions, from low-pressure mixtures to mixtures with preheating at 1 bar. And yet, at more practically relevant atmospheric conditions, they fail, especially in off-stoichiometric compositions.

This immediately raises a number of questions. From the numerical point of view, if the issue is the lack of numerical resolution, then why can the cell size in a lean ($\phi = 0.5$) $H_2$/air mixture be predicted accurately at elevated upstream temperatures, while at room temperature the disagreement is spectacular? This is despite the fact that the ZND induction length is a factor of six larger in the ambient-temperature mixture, which thus requires a much lower numerical resolution.

[4] In the data set by Mikolaitis et al. [154], we include only the $\tau_{ign}$ values corresponding to the first peak in CH*, since it matches the ignition delay obtained from OH* emission.

In the context of the physical description, if the Navier-Stokes model is ill-suited for such extreme flow regimes, then why does it work in certain situations? Furthermore, if the current equilibrium chemistry is the limitation, and explicit treatment of the vibrational non-equilibrium (VNE) must be included in the physicochemical model, as was suggested in the past, then why does the accuracy of the numerical solution depend on the upstream pressure? And why does such accuracy increase with mixture preheating despite the fact that higher post-shock temperatures would shorten ignition delays relative to the vibrational relaxation time thus, at least in principle, creating a more non-equilibrium flow?

It is important to emphasize that we do not imply that VNE is not important in shocked reacting flows. The post-shock state behind a strong shock is certainly in vibrational non-equilibrium. Naïve comparison of the chemical and vibrational time scales, however, may not capture the full complexity of the non-equilibrium reaction kinetics. Without a first-principles reaction model with explicit treatment of the vibrational states of individual species, definitive conclusions cannot be made regarding the importance of VNE in various high-speed combustion regimes.

One of the key limitations in our understanding of high-speed combustion is the lack of robust, direct experimental measurements of the ignition delays, along with the associated speciation data, at high pressures and temperatures, which are representative of the strongly shocked flows. Most importantly, such data must be obtained in realistic mixtures, e.g., in air, rather than in heavily diluted compositions. In particular, we showed above that in hydrogen, existing chemical mechanisms predict qualitatively different dynamics at high pressures $\gtrsim 10$ bar with ignition delays reaching the sub-µs range, resulting in extremely high associated temperature sensitivity. Without systematic, well-characterized, direct measurements in these regimes, it is impossible to establish the accuracy of the existing physicochemical description.

A similar need also exists for hydrocarbon fuels. On one hand, virtually all numerical calculations performed to-date show significant discrepancies between the computed and measured detonation structure. On the other, similar to hydrogen, there is a lack of ignition delay data in the relevant flow regimes. While a substantial amount of data exists for methane/air, other hydrocarbons, such as $C_2H_2$ and especially heavy fuels, are much less explored. For example, for JP-10, existing data suggest a significant discrepancy between predicted and measured ignition delays at high temperatures. This observation must be confirmed by additional independent studies that cover a wider range of conditions.

We also emphasize that experimental studies should not target only a particular pressure and temperature, but rather it is important to explore different paths, through which such a thermodynamic state could form. The same $P$ and $T$ could be produced by a single strong shock or multiple weaker shocks. While existing deterministic equilibrium kinetics does not distinguish between these situations, any processes, which have a history effect, such as vibrational relaxation, would result in a different chemical evolution. Such potential history effects must be explored in experimental studies.

Direct measurements in the extreme strongly shocked thermodynamic regimes are undoubtedly challenging both from the experimental and diagnostic point of view. At the same time, in the absence of such data, it is not possible to rule out thermochemistry as a potential cause of the discrepancies between detonation simulations and experiments. In turn, until the validity of the physical model is firmly established, it is impossible to begin uncovering any other limitations, for instance, in the computational approaches used to study these flows. Together, this creates a conceptual impasse, which inhibits today the advances in the detonation theory.

Finally, it is important for computational studies to shift the focus to more realistic fuels and flow conditions, particularly characterized by more practically relevant pressures. A detonation in a regular mixture at low pressure can inform us about various general properties of this phenomenon. At the same time, we require the ability to make robust predictions of the structure, stability, detonability limits, etc. in various mixtures and under varying conditions. The demonstrated accuracy of such predictions would serve as the ultimate evidence of our understanding of high-speed combustion.

Answering questions raised above requires the concerted effort of the community to match the complexity of the advances needed across a variety of disciplines. Experimental and diagnostic measurements must be adapted to probe ultra-fast chemistry regimes at high pressures and temperatures. Chemical kinetic descriptions may need to move beyond the canonical deterministic finite rate kinetics, and include master-equation-based transition-state models derived from the ab initio calculations. Numerical methods likely may also require novel multi-resolution/multi-fidelity approaches, which go beyond the traditional Navier-Stokes framework. Our hope was to provide, in the limited space available, a brief summary of the challenges that exist in our understanding of high-speed combustion that would motivate such advances.

## CRediT authorship contribution statement

**Alexei Poludnenko**: research conceptualization, wrote and reviewed manuscript, data analysis, funding acquisition. **Sai Sandeep Dammati**: solver development, simulations, data analysis, reviewed and edited manuscript. **Abeetath Ghosh**: simulations, data anal-ysis.

## Declaration of competing interest

The authors declare that they have no known competing financial interests or personal relationships that could have appeared to influence the work reported in this paper.

## Acknowledgments

The authors thank Vadim Gamezo, Brian Taylor, and Daniil Andrienko for valuable discussions. Authors acknowledge funding support by the Air Force Office of Scientific Research award FA9550261B068, and the Office of Naval Research awards N00014-22-1-2606 (Multidisciplinary University Research Initiative) and N00014-23-1-2461. Computing resources were provided by the Department of Defense (DoD) High Performance Computing Modernization Program (HPCMP) and the US Naval Research Laboratory. Authors also gratefully acknowledge assistance of the DoD HPCMP Data Analysis and Visualization (DAV) Center, and in particular of Christopher Lewis, Michael Stephens, Vu Tran, and Miguel Valenciano, with data visualization in Fig. 1.

# Supplementary Material

# State of Understanding of High-speed Combustion: Can the Detonation Structure in Fuel-Air Mixtures Be Predicted?

Alexei Y. Poludnenko[a,b,*], Sai Sandeep Dammati[a,b], Abeetath Ghosh[a]

[a] *School of Mechanical, Aerospace, and Manufacturing Engineering, University of Connecticut, Storrs, CT 06269, USA*

[b] *Center for Clean Energy Engineering, University of Connecticut, Storrs, CT 06269, USA*

Supplementary Table S1: Survey of the numerical studies of detonations in fuel/air (or near-air) mixtures using complex chemistry. Each row represents one set of conditions considered in a given study. $L$ is the largest transverse dimension of a rectangular 3D channel or the tube diameter, $\lambda_c$ is the reported numerical cell width, and $\lambda_{\mathrm{exp}}$ is the experimental cell width for the same mixture. For mixtures listed in Table 2, values of $\lambda_{exp}$ are taken from that table; for other mixtures, values of $\lambda_{exp}$ are taken from the corresponding studies, if such values are provided. For studies that consider several grid resolutions, only the finest one is listed.

| **Ref.** | **Mixture** | **$P$ (bar)** | **$T$ (K)** | **$\phi$** | **NS / Euler** | **Geometry** | **Channel size, $L$ (cm)** | **Soot foil provided?** | **$\lambda_c$ (cm)** | **$L/\lambda_c$** | **$\lambda_c/\lambda_{\mathrm{exp}}$** |
|---|---|---|---|---|---|---|---|---|---|---|---|
| **$H_2$/air** | **sub-atmospheric: $P$ < 1 bar** | | | | | | | | | | |
| [1] | $H_2$/air | 0.13 | 293 | 1 | Euler | 2D | 2.35 | Yes | 2.4 [a] | 0.98 | – |
| " | $H_2$/air | 0.43 | 293 | 1 | Euler | 2D | 1.52 | Yes | 1.5 [a] | 1.01 | – |
| [2] | $H_2$/air | 0.25 | 300 | 1 | Euler | 2D | 5.53 | No[b] | 1.08 | 5.12 | 0.45 |
| " | $H_2$/air | 0.25 | 300 | 1 | NS | 2D | 5.53 | No[b] | 1.09 | 5.07 | 0.45 |
| " | $H_2$/air | 0.25 | 300 | 1 | NS | 2D | 5.53 | No[b] | 0.75 | 7.37 | 0.31 |
| " | $H_2$/air | 0.25 | 300 | 1 | Euler | 2D | 3.77 | No[b] | 0.46 | 8.20 | 0.19 |
| " | $H_2$/air | 0.25 | 300 | 1 | NS | 2D | 3.77 | No[b] | 0.45 | 8.38 | 0.19 |
| [3] | $H_2$/air | 0.40 | 300 | 1 | NS | 2D | 5.79 | Yes | 1.19 | 4.87 | 0.91 |
| " | $H_2$/air + $O_3$ | 0.40 | 300 | 1 | NS | 2D | 2.5 | Yes | 0.42 | 5.95 | 0.67 |
| **$H_2$/air** | **near-ambient: $P$ = 1 bar, $T$ = 273–300 K** | | | | | | | | | | |
| [1] | $H_2$/air | 1.0 | 293 | 1 | Euler | 2D | 0.67 | Yes | 0.67 [a] | 1.0 | 0.55 |
| " | $H_2$/air | 1.0 | 293 | 1 | Euler | 2D | 0.084 | Yes | 0.084 [a] | 1.0 | 0.069 |
| [4] | $H_2$/air | 1.0 | 273 | 1 | Euler [h] | 2D | 0.2 | Yes | 0.2 [a] | 1.0 | 0.17 |
| " | $H_2$/air | 1.0 | 273 | 1 | Euler [h] | 2D | 0.2 | Yes | 0.1 [a] | 2.0 | 0.08 |
| [5] | $H_2$/air | 1.0 | N/A [c] | N/A [c] | Euler | 2D | 0.05 | Yes | 0.05 [a] | 1.00 | 0.04 |
| " | $H_2$/air | 1.0 | N/A [c] | N/A [c] | Euler | 3D square | 0.05 | Yes | 0.05 [a] | 1.00 | 0.04 |
| [6] | $H_2$/air | 1.0 | 300 | 1 | Euler | 3D square | 0.1 | No | N/A[d] | N/A | N/A |
| [7] | $H_2$/air | 1.0 | 300 | 1 | Euler | 3D tube | 0.1 | Yes | N/A [e] | < 1 [e] | N/A |
| " | $H_2$/air | 1.0 | 300 | 1 | Euler | 3D square | 0.1 | Yes | N/A [e] | < 1 [e] | N/A |
| [8] | $H_2$/air | 1.0 | 300 | 1 | Euler | 3D square | 0.1 | Yes | N/A [e] | < 1[e] | N/A |
| [9] | $H_2$/air | 1.0 | 300 | 1 | Euler | 2D | 0.2 | Yes | 0.3 [a] | 0.67 | 0.25 |
| " | $H_2$/air | 1.0 | 300 | 1 | Euler | 3D tube | 0.1 | Yes | N/A [e] | < 1 [e] | N/A |
| [10] | $H_2$/air | 1.0 | 300 | 1 | Euler | 2D | 4.0 | Yes | 0.1–0.2 | 20.0–40.0 | 0.08–0.17 |
| [11] | $H_2$/air | 1.0 | 293 | 1 | Euler | 2D | 4.0 | Yes | 0.3–0.4 | 10.0–13.33 | 0.25–0.33 |
| [12] | $H_2$/air [f] | 1.0 | 300 | 1 | Euler | 2D | 5.0 | Yes | 0.2 | 25.0 | 0.17 |
| [13] | $H_2$/air | 1.0 | 298 | 1 | Euler [g] | 2D | 1.38 | Yes | 0.355 | 3.89 | 0.29 |
| " | $H_2$/air | 1.0 | 298 | 1 | Euler [g] | 2D | 3.31 | Yes | 0.740 | 4.47 | 0.61 |
| **$H_2$/air** | **preheated: $P$ = 1 bar, $T$ =400–650 K** | | | | | | | | | | |
| [14] | $H_2$/air [f] | 1.0 | 500 | 1 | Euler [h] | 2D | 5.0 | No | 0.67 | 7.46 | 1.06 |
| " | $H_2$/air [f] | 1.0 | 500 | 1 | Euler [h] | 2D | 5.0 | No | 0.44 | 11.36 | 0.70 |
| " | $H_2$/air [f] | 1.0 | 500 | 1 | Euler [h] | 2D | 5.0 | No | 0.40 | 12.5 | 0.63 |
| " | $H_2$/air [f] | 1.0 | 500 | 1 | Euler [h] | 2D | 5.0 | No | 0.36 | 13.89 | 0.57 |
| " | $H_2$/air [f] | 1.0 | 500 | 1 | Euler [h] | 2D | 5.0 | No | 0.33 | 15.15 | 0.52 |
| " | $H_2$/air [f] | 1.0 | 500 | 1 | Euler [h] | 2D | 5.0 | No | 0.31 | 16.13 | 0.49 |
| [12] | $H_2$/air [f] | 1.0 | 400 | 1 | Euler | 2D | 5.0 | Yes | 0.4 | 12.5 | – [i] |
| " | $H_2$/air [f] | 1.0 | 450 | 1 | Euler | 2D | 5.0 | Yes | 0.56 | 8.93 | – [i] |
| " | $H_2$/air [f] | 1.0 | 500 | 1 | Euler [h] | 2D | 5.0 | Yes | 0.7 | 7.14 | 1.11 |
| " | $H_2$/air [f] | 1.0 | 500 | 1 | Euler [h] | 2D | 5.0 | Yes | 0.67 | 7.46 | 1.06 |
| " | $H_2$/air [f] | 1.0 | 500 | 1 | Euler [h] | 2D | 5.0 | Yes | 0.44 | 11.36 | 0.70 |
| " | $H_2$/air [f] | 1.0 | 500 | 1 | Euler [h] | 2D | 5.0 | Yes | 0.40 | 12.5 | 0.63 |
| " | $H_2$/air [f] | 1.0 | 500 | 1 | Euler [h] | 2D | 5.0 | Yes | 0.36 | 13.89 | 0.57 |
| " | $H_2$/air [f] | 1.0 | 500 | 1 | Euler [h] | 2D | 5.0 | Yes | 0.33 | 15.15 | 0.52 |
| " | $H_2$/air [f] | 1.0 | 500 | 1 | Euler [h] | 2D | 5.0 | Yes | 0.31 | 16.13 | 0.49 |
| " | $H_2$/air [f] | 1.0 | 650 | 1 | Euler | 2D | 5.0 | Yes | 0.5 | 10.0 | 1.11 |

*continued on next page*

Supplementary Table S1 – *continued*

| **Ref.** | **Mixture** | $P$ **(bar)** | $T$ **(K)** | $\phi$ | **NS / Euler** | **Geometry** | **Channel size,** $L$ **(cm)** | **Soot foil provided?** | $\lambda_c$ **(cm)** | $L/\lambda_c$ | $\lambda_c/\lambda_{exp}$ |
|---|---|---|---|---|---|---|---|---|---|---|---|
| **$H_2$/air** | **high pressure and temperature: $P \approx 5$ bar, $T$ = 950 K** | | | | | | | | | | |
| [15] | $H_2$/air | 5.07 | 950 | 1 | NS | 2D | 0.8 | Yes | 0.055 | 14.55 | – [i] |
| " | $H_2$/air | 5.07 | 950 | 1 | NS [j] | 2D | 0.8 | Yes | 0.055 | 14.55 | – [i] |
| [16] | $H_2$/air | 5.07 | 950 | 1 | NS | 2D | 0.1 | Yes | 0.1 [a] | 1.00 | – [i] |
| [17] | $H_2$/air | 5.07 | 950 | 1 | NS | 3D square | 0.1 | No | 0.05 [a] | 2.00 | – [i] |
| [18] | $H_2$/air | 5.07 | 950 | 1 | NS | 2D | 0.11 | Yes | 0.055 [a] | 2.00 | – [i] |
| " | $H_2$/air | 5.07 | 950 | 1 | NS | 2D | 0.22 | Yes | 0.055 | 4.00 | – [i] |
| **$CH_4/O_2/N_2$** | **$P$ = 1 bar, $T$ = 298 K** | | | | | | | | | | |
| [19] | $CH_4/O_2/N_2$ [k] | 1.0 | 298 | 1 | NS | 2D | 10.0 | No | 0.9 | 11.11 | 0.41 [l] |
| " | $CH_4/O_2/N_2$ [k] | 1.0 | 298 | 1 | NS | 2D | 6.0 | No | 1.75 | 3.43 | 1.40 [m] |
| [20] | $CH_4/O_2/N_2$ [k] | 1.0 | 298 | 1 | NS | 2D | 10.0 | Yes | 1.49 [n] | 6.71 | – [i] |
| " | $CH_4/O_2/N_2$ [k] | 1.0 | 298 | 1 | NS | 2D | 6.0 | Yes | 1.22 [n] | 4.92 | 1.14 |
| **$C_2H_2/O_2/N_2$** | **$P$ = 1 bar, $T$ = 298 K** | | | | | | | | | | |
| [19] | $C_2H_2/O_2/N_2$ [k] | 1.0 | 298 | 0.4 | NS | 2D | 2.0 | No | 0.52 | 3.85 | 0.95 [m] |
| [16] | $C_2H_2/O_2/N_2$ [k] | 1.0 | 298 | 0.4 | NS | 2D | 2.0 | Yes | 0.5 [m,o] | 4.00 | 0.91 [m] |
| [20] | $C_2H_2/O_2/N_2$ [k] | 1.0 | 298 | 0.4 | NS | 2D | 2.0 | Yes | 0.96 [a,n] | 2.08 | 0.70 |
| **JP-10/air** | | | | | | | | | | | |
| [21] | JP-10/air | 1.0 | 373 [p] | 1 | Euler | 2D | 25.0 | Yes | 5.0 [o] | 5.0 | 0.96 |

[a] The cell width is close to or smaller than the channel size.
[b] The cells are shown as maps of $D/D_{CJ}$.
[c] The initial temperature and the equivalence ratio are not reported.
[d] No cell width reported.
[e] Spinning detonation.
[f] The composition is close to, but not exactly, air.
[g] Modified chemistry models based on the uncertainties of the reaction rate.
[h] Simulations use different chemical kinetic mechanisms.
[i] Experimental measurements not available.
[j] Chemical kinetics with modified termolecular reactions.
[k] $O_2$-enriched air.
[l] $\lambda_{exp}$ is obtained from the literature. The UCF experiment for this mixture failed to detonate.
[m] $\lambda_{exp}$ is from thin-channel UCF experiments reported in [20, 22].
[n] Cell length is reported, rather than cell width.
[o] Detonation has not reached a steady state or no temporal convergence of the cellular structure.
[p] Simulations performed at 373 K; experiments used for comparison performed at 353 K.

Supplementary Table S2: Survey of the experimental studies of the detonation cell sizes in fuel-air mixtures, grouped by the mixture and its initial conditions ($P$, $T$, $\phi$). Within each mixture block, each row represents a separate experimental study. "H/D" is the largest transverse dimension in a rectangular channel or the tube inner diameter; "W" is the narrowest transverse dimension in a rectangular channel. The final column is the channel or tube size divided by the average reported cell size. Mixtures 7 and 11 in Table 2 of the main manuscript are not included, since their values of $\lambda_{exp}$ were obtained, respectively, via extrapolation of the existing measurements and from a phenomenological correlation (see Table 2).

| Ref. | Geometry | H or D (cm) | W (cm) | Reported detonation speed | Cell size, $\lambda_{exp}$ (cm) | Measurement uncertainty (cm) | Channel / Cell size |
|---|---|---|---|---|---|---|---|
| **$H_2$/air** $P$ = 1 **bar,** $T$ = 298 **K,** $\phi$ = 1 | | | | | | | |
| [23] | Tube | 10.0 | – | Within 2% of $D_{CJ}$ (+) | 1.0 | N/A | 10.0 |
| [24] | Tube | 27.0 | – | Within 2% of $D_{CJ}$ (+) | 0.8 | 0.2 | 33.75 |
| [25] | Tube | 15.0 | – | No comment made | 1.5 | N/A | 10.0 |
| [26] | Tube | 20.0 | – | Matches CJ speed | 1.0, 1.55 | N/A | 20.0, 12.9 |
| [27] | Tube | 43.0 | – | Matches CJ speed | 0.99 | 0.18 | 43.43 |
| [28] | Tube | 43.0 | – | Steady state claimed | 1.25 | 0.2 | 34.4 |
| [29] | Channel | 7.6 | 3.8 | Within 3% of $D_{CJ}$ (−) | 1.59 | 0.19 | 4.78, 2.39[a] |
| **$H_2$/air** $P$ = 0.25 **bar,** $T$ = 293 **K,** $\phi$ = 1 | | | | | | | |
| [28] | Tube | 43.0 | – | $D$ typically within a few % of CJ (no specific data) | 2.4 | Factor of 2 scatter [b] | 17.92 |
| **$H_2$/air** $P$ = 1 **bar,** $T$ = 650 **K,** $\phi$ = 1 | | | | | | | |
| [24] | Tube | 27.0 | – | Within 2–3% of $D_{CJ}$ (±) | 0.45 | 0.05 | 60.0 |
| [23] | Tube | 10.0 | – | Within 2% of $D_{CJ}$ (+) | 0.45 | 0.05 | 22.22 |
| [30] | Tube | 10.0 | – | Within 2–3% of $D_{CJ}$ (±) | 0.45 | 0.05 | 22.22 |
| **$H_2$/air** $P$ = 1.01 **bar,** $T$ = 298 **K,** $\phi$ = 0.5 | | | | | | | |
| [31] | Tube | 43.0 | – | Within 2% of $D_{CJ}$ (±) | 10.15 | 25% of mean | 4.24 |
| [32] | Tube | 43.0 | – | Within 2% of $D_{CJ}$ (±) | 10.15 | 25% of mean | 4.24 |
| **$H_2$/air** $P$ = 1.01 **bar,** $T$ = 650 **K,** $\phi$ = 0.5 | | | | | | | |
| [24] | Tube | 27.0 | – | Within 2% of $D_{CJ}$ (+) | 1.92 | 0.48 | 14.06 |
| **$CH_4$/air** $P$ = 1.01 **bar,** $T$ = 298 **K,** $\phi$ = 1 | | | | | | | |
| [33] | Tube | 183.0 | – | Claims near-CJ propagation (no supporting evidence) | 28.0 | N/A | 6.54 |
| [34][‡] | Tube | 105.0 | – | 0.2% of $D_{CJ}$ reported | 25.0 | 30% of mean | 4.2 |
| **$C_2H_4$/air** $P$ = 1 **bar,** $T$ = 298 **K,** $\phi$ = 0.7 | | | | | | | |
| [35][¶] | Tube | 89.0 | – | Within 5% of $D_{CJ}$ (±) | 8.0 | 12.5% of mean | 11.13 |
| [36] | Tube | 89.0 | – | Claims near-CJ propagation (no supporting evidence) | 8.0 | 12% of mean | 11.13 |
| **$C_2H_2$/air** $P$ = 1.01 **bar,** $T$ = 293 **K,** $\phi$ = 3 | | | | | | | |
| [37] | Tube | 5.0 | – | No comment made [c] | 1.16 | N/A | 4.31 |
| **JP-10/air** $P$ = 1 **bar,** $T$ = 353 **K,** $\phi$ = 1.4 | | | | | | | |
| [38] | Tube | 28.0 | – | Within 1% of $D_{CJ}$ (±) | 7.41 | 4.0 min, 9.9 max | 3.78 |

[‡] Natural gas, not pure methane.
[¶] $P$ = 93.2 kPa.
[a] Ratio of the largest and smallest transverse dimensions of the rectangular channel to the reported cell size. The narrowest channel size $2.39\lambda_{exp}$ is close to $\lambda_{exp}$, however the detonation speed was reported to be within 3% of $D_{CJ}$, which suggests a minimal effect of wall losses.
[b] A general "factor of 2 scatter" comment was made for the entire dataset.
[c] Although no detonation speed is reported, the tube diameter of $4.31\lambda_{exp}$ is sufficiently larger than $\lambda_{exp}$ to suggest a minimal effect of wall losses.

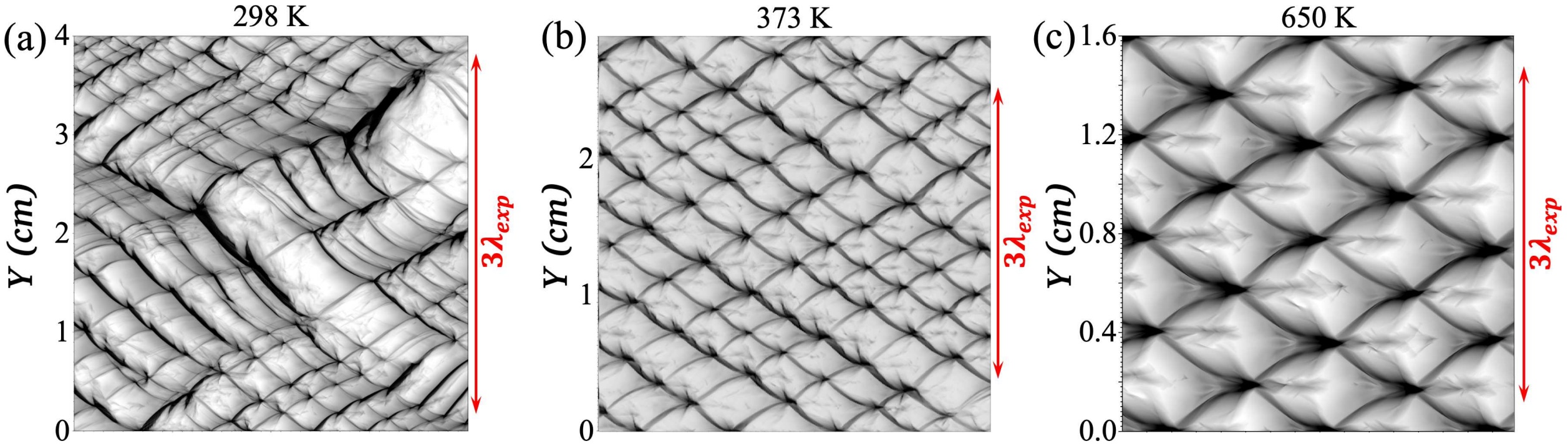

Figure S1: Synthetic soot foils in the stoichiometric $H_2$/air mixtures at 1 bar. Shown is the logarithm of maximum pressure recorded at each point. (a) 298 K (mixture 1); (b) 373 K; (c) 650 K (mixture 3). Domain width relative to the corresponding experimental cell size $\lambda_{exp}$ is shown in each panel. The average experimental cell size in the mixture in panel (b) is $\lambda_{exp} = 0.7 \pm 0.18$ cm [27]. The experimental value of $\lambda_{exp}$ for this mixture satisfies the selection criteria described in § 2. In all panels, detonation propagates from right to left.

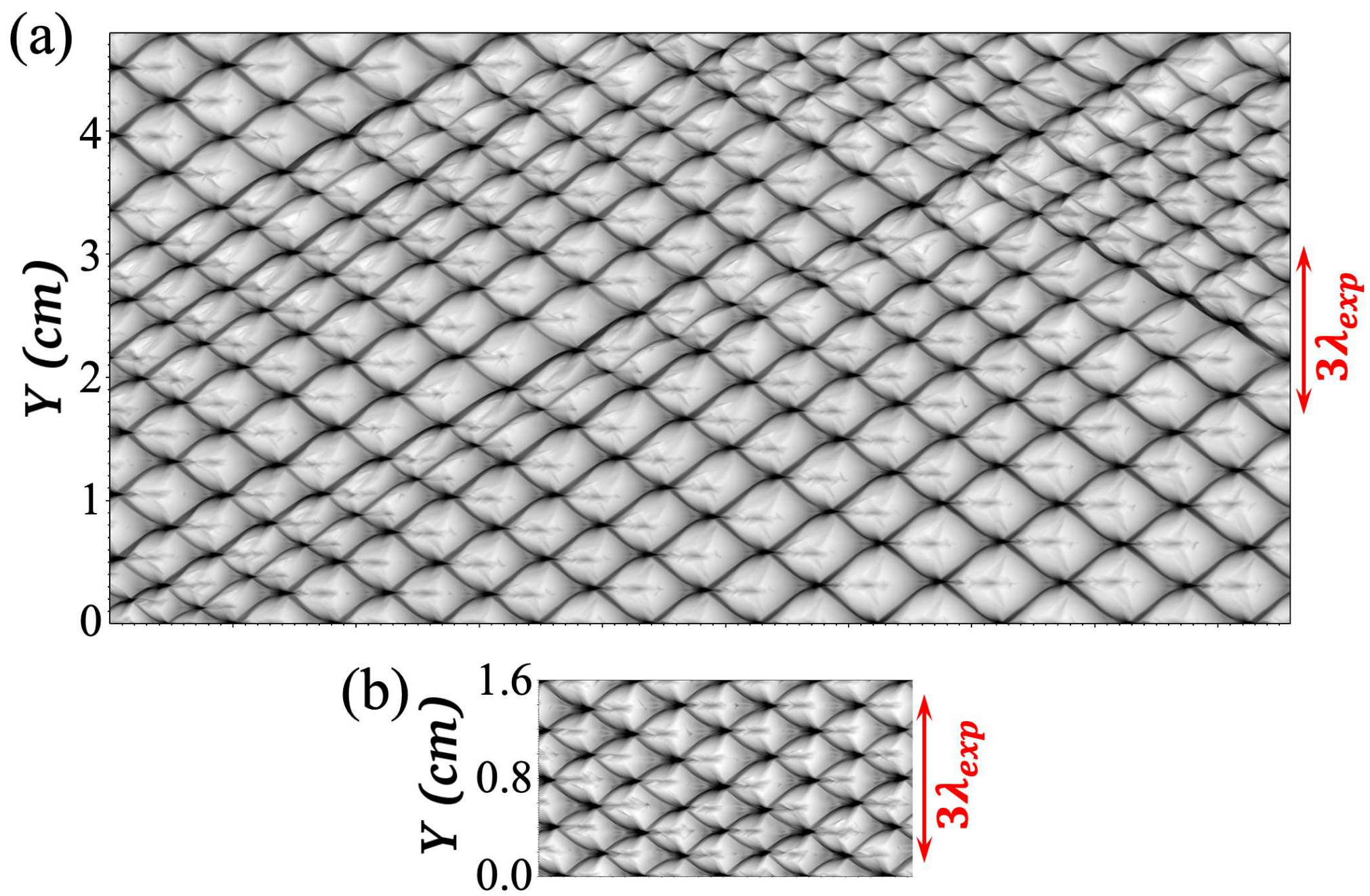

Figure S2: Numerical soot foils in a stoichiometric $H_2$/air at 1 bar, 650 K (mixture 3). Shown is the logarithm of maximum pressure recorded at each point. (a) The 2D calculation performed in a 4.8 cm wide domain compared to a 1.6 cm wide domain in the calculation shown in Figs. 2a and 3a,b. (b) Soot foil in Fig. 2a shown for comparison on the same scale as panel (a).

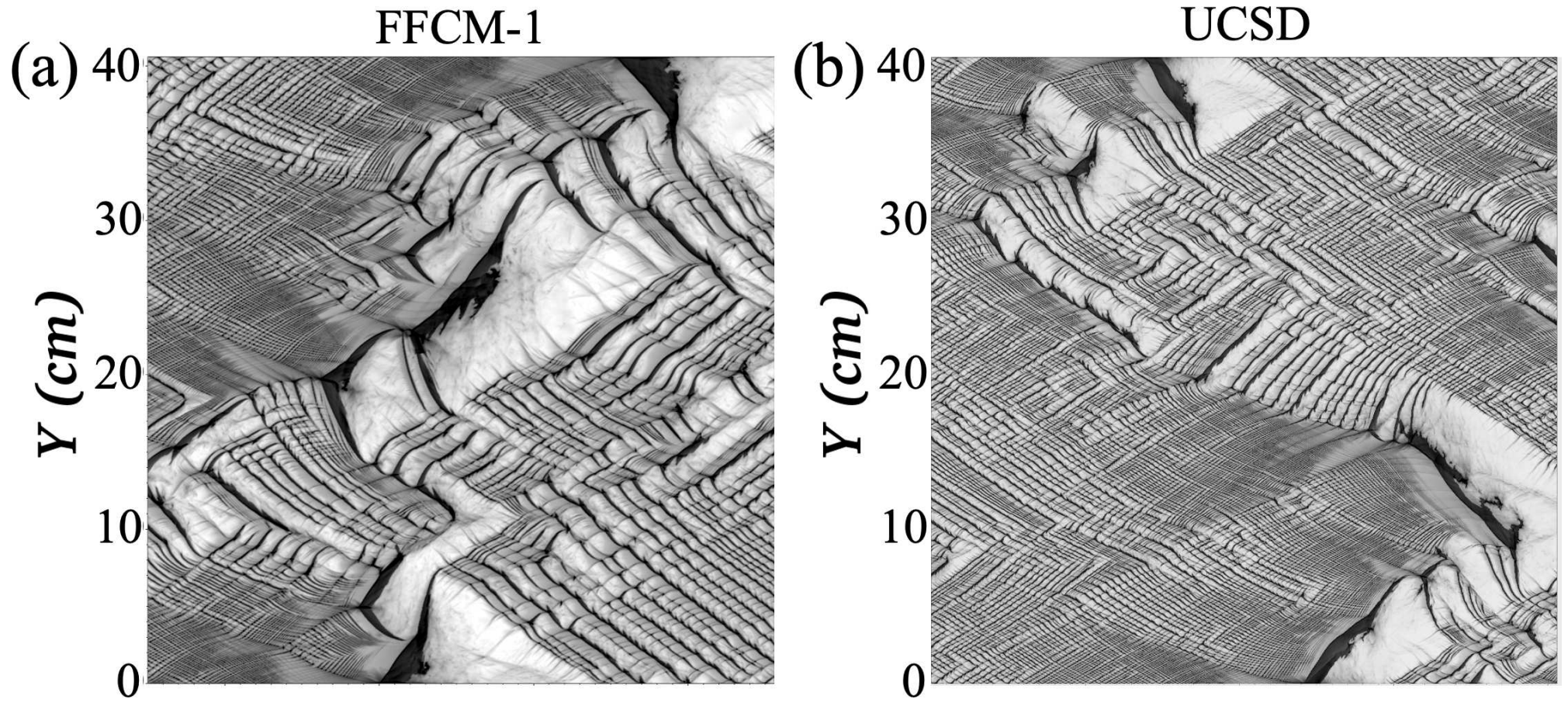

Figure S3: Numerical soot foil in $H_2$/air at $\phi$ = 0.5 and 1 bar, 298 K (mixture 4). Shown is the logarithm of maximum pressure recorded at each point in the 2D simulations performed using the FFCM-1 [39] (a) and the UCSD chemical mechanisms [40] (b).

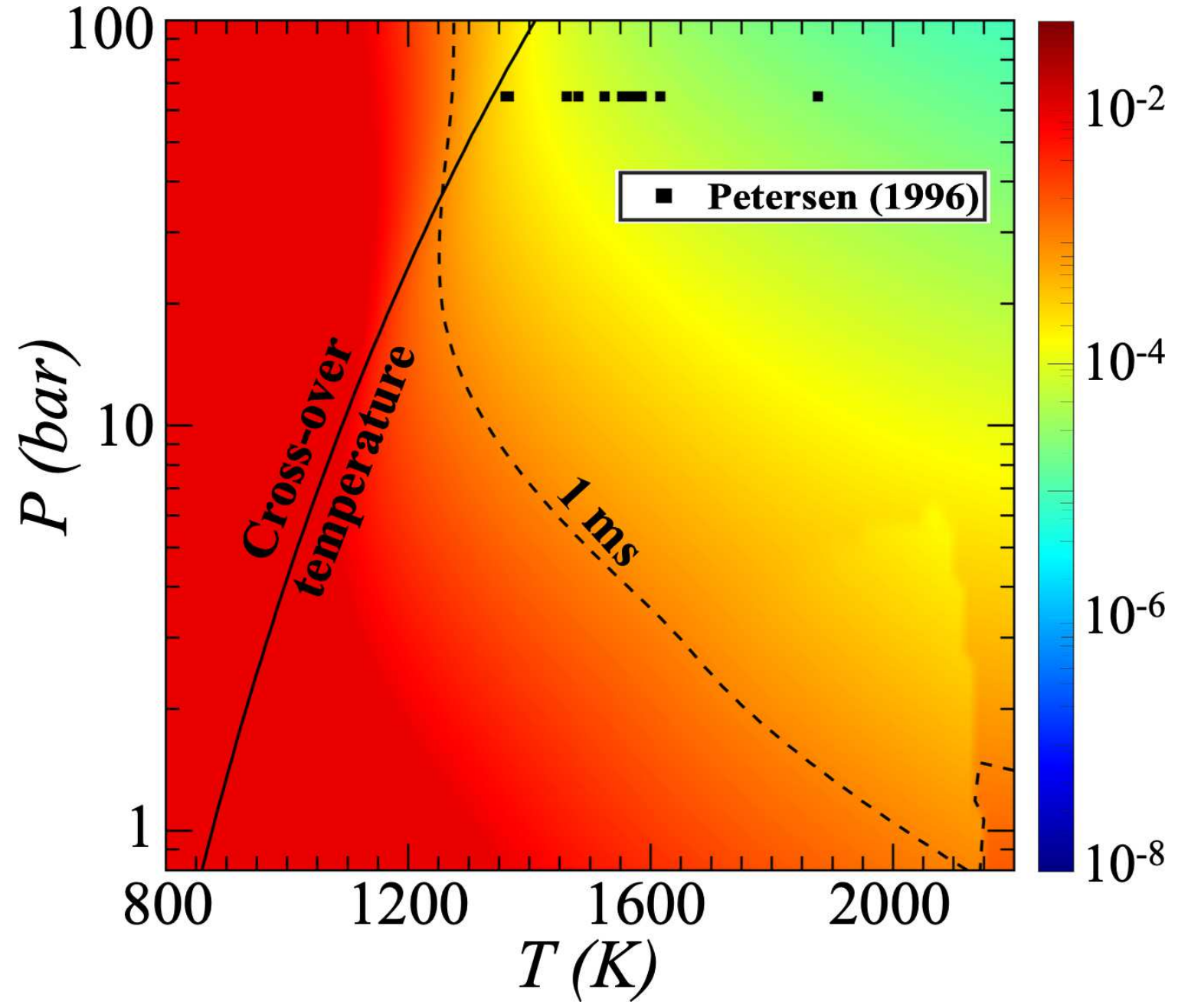

Figure S4: Ignition times, $\tau_{ign}$ (s), in the mixture with 0.1%$H_2$, 0.05%$O_2$, and 99.85%Ar (by mole) used in the experimental measurements of $\tau_{ign}$ by Petersen *et al.* [41] (square symbols). Values of $\tau_{ign}$ were calculated using the FFCM-1 [39] chemical mechanism.